\input harvmac
%\draftmode
%
\rightline{WIS/97/9, EFI-97-20, RI-2-97}
\Title{
\rightline{hep-th/9704104}
}
{\vbox{\centerline{Brane Dynamics and N=1 Supersymmetric Gauge Theory}}}
\medskip

\centerline{\it S. Elitzur${}^1$, 
A. Giveon${}^1$,  D. Kutasov${}^{2,}$\footnote{*}{On leave of absence
from Department of Physics, University of Chicago,
5640 S. Ellis Ave., Chicago, IL 60637, USA.}, 
E. Rabinovici${}^1$, A. Schwimmer${}^{3,4}$}
\bigskip
\centerline{${}^1$Racah Institute of Physics, The Hebrew University,
Jerusalem, Israel}

\centerline{${}^2$Department of Physics of Elementary Particles,
Weizmann Institute of Science, Rehovot, Israel}

\centerline{${}^3$Department of Physics of Complex Systems,
Weizmann Institute of Science, Rehovot, Israel}

\centerline{${}^4$Theory Division, CERN, Geneva, Switzerland}
\smallskip

\vglue .3cm
%\vskip 2cm
\bigskip

\noindent
We discuss some aspects of the relation between 
space-time properties of branes in string theory, 
and the gauge theory on their worldvolume, for 
models invariant under four supercharges in three 
and four dimensions. We show that a simple set of
rules governing brane dynamics reproduces many 
features of gauge theory. We study theories 
with $U(N_c)$, $SO(N_c)$ and $Sp(N_c)$ gauge groups 
and matter in the fundamental and two index tensor 
representations, and use the brane description to 
establish Seiberg's electric-magnetic duality for 
these models.

\Date{4/97}

\def\journal#1&#2(#3){\unskip, \sl #1\ \bf #2 \rm(19#3) }
\def\andjournal#1&#2(#3){\sl #1~\bf #2 \rm (19#3) }

\def\ie{{\it i.e.}}
\def\eg{{\it e.g.}}

\def\frac#1#2{{#1\over#2}}

\def\inbar{\,\vrule height1.5ex width.4pt depth0pt}
\def\IC{\relax\hbox{$\inbar\kern-.3em{\rm C}$}}
\def\IR{\relax{\rm I\kern-.18em R}}
\def\IP{\relax{\rm I\kern-.18em P}}

%
%%%%%%%%%%%%%%%%%%%%%%%%%%%%%%%%%%%%
%
\def\np#1#2#3{Nucl. Phys. {\bf B#1} (#2) #3}
\def\pl#1#2#3{Phys. Lett. {\bf #1B} (#2) #3}

\def\prl#1#2#3{Phys. Rev. Lett. {\bf #1} (#2) #3}

\def\ap#1#2#3{Ann. Phys. {\bf #1} (#2) #3}

\catcode`\@=11
\def\slash#1{\mathord{\mathpalette\c@ncel{#1}}}
\overfullrule=0pt

\def\underrel#1\over#2{\mathrel{\mathop{\kern\z@#1}\limits_{#2}}}

\catcode`\@=12

%%%%%%%%%%%%%%%%%%%%%%%%%%%%%%%%%%%%%%%%%%%%%%%%%%%%%%%%%%%%%%

%

\def\det{{\rm det}}

\def\det{{\rm det}}

%%%%%%%%%%%%%%%%%%%%%%%%%%%%%%%%%%%%%%%%%%%%%%%%%%%%%%%%%%%%%%
% new defs:

\def\nsp{{NS$^\prime$}}

%%%%%%%%%%%%%%%%%%%%%%%%%%%%%%%%%%%%%%%%%%%%%%%%%%%

\newsec{Introduction}

In the past three years there has been remarkable
progress in field and string theory. In four
dimensional field theory with $N\ge 1$ supersymmetry,
a combination of symmetries and holomorphicity of
certain terms in the low energy quantum effective
action was found to provide constraints
that in many cases are sufficient to completely
determine the strongly coupled low energy dynamics.
These constraints become more powerful as the number of
supersymmetries increases. Theories with
fewer supersymmetries typically have richer dynamics,
but the handle one has on their dynamics is more limited.

\nref\mo{C. Montonen and D. Olive, \pl{72}{1977}{117}.}%
\nref\sw{N. Seiberg and E. Witten, hep-th/9407087,
\np{426}{1994}{19}; hep-th/9408099, \np{431}{1994}{484}.}%
\nref\neqone{N. Seiberg, hep-th/9408013;
hep-th/9506077.}%
\nref\nonerev{K. Intriligator and N. Seiberg,
hep-th/9509066.}%
The organizing
principle that seems to emerge is electric-magnetic,
strong-weak coupling duality.
$N=4$ supersymmetric gauge theories
are now believed to exhibit an exact
Montonen-Olive duality \mo.
$N=2$ supersymmetric theories,
as discussed by Seiberg and Witten
\sw, can be solved
(in the BPS sector) using duality and supersymmetry.
$N=1$ supersymmetric gauge theories have the richest
dynamics of the three,
are the closest to phenomenology, and in
many cases can be studied at low energies by
a generalization of Montonen-Olive duality due
to Seiberg \refs{\neqone, \nonerev}. However,
while there are many examples of $N=1$ supersymmetric
gauge theories that can be analyzed
using duality, the situation is not as satisfactory as that
for theories with $N>1$ supersymmetry,
since there are many cases for which
the standard techniques appear to be inadequate.
$N=1$ theories that are understood,  are usually analyzed
on a case by case basis with no obvious unified picture.
There is no general apriori understanding of whether any
given model possesses a useful dual.

\nref\doug{M. Douglas, hep-th/9604198.}%
\nref\sen{A. Sen, hep-th/9605150, \np{475}{1996}{562}.}%
\nref\bds{T. Banks, M. Douglas and N. Seiberg, hep-th/9605199,
\pl{387}{1996}{278}.}%
\nref\sone{N. Seiberg, hep-th/9606017,
\pl{384}{1996}{81};
hep-th/9608111, \pl{388}{1996}{753}.}%
\nref\hw{A. Hanany and E. Witten, hep-th/9611230.}%
\nref\is{K. Intriligator and N. Seiberg,
hep-th/9607207, \pl{387}{1996}{513}.}%
\nref\ooo{J. de Boer, K. Hori, H. Ooguri and Y. Oz,
hep-th/9611063;
J. de Boer, K. Hori, H. Ooguri, Y. Oz and
Z. Yin, hep-th/9612131.}%
\nref\pz{M. Porrati and A. Zaffaroni, hep-th/9611201.}%
One of the most interesting recent developments is the realization
that many of the field theory phenomena mentioned
above can be understood by embedding gauge theories in
string theory and using properties of the latter to study the former.
A very useful idea has been to use the interplay
between the geometry and dynamics of branes,
and the quantum field theory on their worldvolume
\refs{\doug - \sone}. In particular, in \hw\ it was shown
that applying string duality to a configuration of branes
in type IIB string theory provides an explanation of a certain
``mirror symmetry'' in 2+1 dimensional $N=4$ supersymmetric 
gauge theory discovered in \is\ (see also \refs{\ooo, \pz}).

\nref\egk{S. Elitzur, A. Giveon and D. Kutasov,
hep-th/9702014.}%
\nref\kms{D. Kastor, E. Martinec and S. Shenker,
\np{316}{1989}{590}.}%
\nref\martinec{E. Martinec, \pl{217}{1989}{431}.}%
\nref\vw{B. Greene, C. Vafa and N. Warner, 
\np{324}{1989}{371}.}%
\nref\ew{E. Witten, hep-th/9301042,
\np{403}{1993}{159}.}%
In a recent paper \egk\
it has been proposed that configurations of branes in type
II string theory preserving four supercharges provide a natural
arena for the study of four dimensional $N=1$ supersymmetric gauge
dynamics. Seiberg's duality relates theories
that, in the brane construction, can be connected by
a continuous path in the  moduli space of vacua, along which
the gauge symmetry is completely broken
and the infrared dynamics is weak. The relation between the electric and
magnetic theories in this framework is reminiscent of the well known
correspondence between two dimensional sigma models on Calabi-Yau
hypersurfaces in weighted projective spaces and Landau-Ginzburg
models with $N=(2,2)$ SUSY \refs{\kms-\vw}; the continuous
path is analogous to the interpolation between the CY and LG
descriptions of the theory, both of which are embedded in the larger
framework of the (non-conformal) gauged linear sigma model \ew.

The purpose of this paper is to
further study the dynamics of branes
realizing four dimensional $N=1$
supersymmetric field theories on their
worldvolume. The strategy is to use the
interplay between brane dynamics and gauge
theory to shed light on both. One can use
simple gauge theory phenomena to learn about
properties of branes, and then use these
properties to study more complicated gauge
theory phenomena using branes. We will use
these ideas to complete and extend the explanation
of Seiberg's duality from string theory proposed
in \egk, and present some related results
in three and four dimensional gauge theories.
As will become clear below, our discussion
will touch on only a small subset of the many
possible questions. A better understanding of
the brane-gauge theory dictionary is likely to
improve both our understanding of the role of
branes in string theory and of gauge theory dynamics.

\nref\vf{C. Vafa, hep-th/9602022,
\np{469}{1996}{403}.}%
\nref\vvv{M. Bershadsky, A. Johansen,
T. Pantev, V. Sadov and C. Vafa,
hep-th/9612052; C. Vafa and
B. Zwiebach, hep-th/9701015.}%
\nref\ov{H. Ooguri and C. Vafa, 
hep-th/9702180.}
Another approach to studying the
dynamics of $N=1$ supersymmetric
four dimensional gauge theory and 
in particular Seiberg's duality 
that has been discussed recently 
is F theory \refs{\vf-\ov}. It 
provides a complimentary picture 
of gauge dynamics by a different
geometrization of the gauge theory
phenomena. Our approach and
F theory are related by T-duality.
It is clearly important to pursue
both approaches further to gain
insight into strongly coupled gauge
theory and string theory.

The plan of this paper is as follows.
In sections 2 -- 4 we introduce the set
of branes and orientifolds which we use
in subsequent sections to construct
different gauge theories. We check
that these branes preserve $N=1$ SUSY
in four dimensions, and describe
some of their classical and quantum
properties. These properties can be
deduced by constructing gauge theories
out of brane configurations, and requiring
that brane dynamics reproduce known
properties of the gauge theory. This
approach to gauge/brane dynamics provides
a nice geometric realization of the moduli
space of vacua in gauge theory, and of the
space of deformations. In the course of the
discussion we often find it convenient
to compactify one of the four spacetime
coordinates on a circle of radius $R$ and
study the physics as a function of $R$.
This leads to interesting connections to recent
discussions of three dimensional $N=2$
supersymmetric gauge theories.

In sections 5, 6 we discuss four dimensional
$N=1$ supersymmetric QCD and its compactification
on a circle. We start in section 5
by constructing a configuration of branes
describing at low energies $N=1$ supersymmetric
Yang-Mills theory with gauge group $U(N_c)$
and $N_f$ flavors of quarks in the
fundamental representation (we will
refer to this as the ``electric'' theory).
We describe the classical space of
deformations of the model
(the moduli space of vacua and
mass deformations) and discuss
the global symmetries and their realization
in the brane configuration.
We also
describe a different brane configuration
corresponding to a ``magnetic''
SQCD theory with gauge group $U(N_c)$,
$N_f$ flavors of quarks and a gauge
singlet ``magnetic meson'' field which
couples to the quarks via a cubic superpotential.
We show that the Higgs moduli space of the
electric theory with $N_c$ colors and
$N_f$ flavors and that of the magnetic
theory with $N_f-N_c$ colors and $N_f$
flavors provide different descriptions of
a single smooth space -- the moduli
space of supersymmetric brane configurations.
This is the string theory origin of the
relation between the two theories,
discovered in gauge theory by Seiberg
\ref\nati{N. Seiberg, hep-th/9411149,
\np{435}{1995}{129}.}.

In section 6 we turn to quantum effects.
Using the results of section 4 we
reproduce the known quantum global
symmetries, moduli spaces and deformations
of SQCD with different numbers of
flavors $N_f$, in three and four dimensions.
We discuss quantum effects in the ``magnetic''
SQCD model of section 5, and complete the proof
of Seiberg's duality from string theory.

In section 7 we consider $U(N_c)$
gauge theory with one or two adjoint
superfields with polynomial
superpotentials, as well as a number
of fundamentals. We recover the field
theory results of
\ref\ks{D. Kutasov, hep-th/9503086,
\pl{351}{1995}{230};
D. Kutasov and A. Schwimmer,
hep-th/9505004, \pl{354}
{1995}{315}; D. Kutasov, A. Schwimmer
and N. Seiberg, hep-th/9510222,
\np{459}{1996}{455}.},
and study some generalizations of that work.

\nref\iss{K. Intriligator and
N. Seiberg, hep-th/9503179,
\np{444}{1995}{125}.}%
\nref\ip{K. Intriligator and
P. Pouliot, hep-th/9505006,
\pl{353}{1995}{471}.}%
\nref\intrrr{ K. Intriligator,
hep-th/9505051,
\np{448}{1995}{187}.}%
\nref\ls{ R.G. Leigh and M.J. Strassler,
hep-th/9505088,
\pl{356}{1995}{492}.}%
\nref\ils{K. Intriligator,
R. Leigh and M. Strassler,
hep-th/9506148, \np{456}{1995}{567}.}
In sections 8, 9 we discuss the case
of symplectic and orthogonal groups,
with matter in the fundamental, symmetric
and antisymmetric tensor representations.
To study such theories we introduce,
in addition to branes, orientifold
hyperplanes. We reproduce many of
the results obtained in gauge theory
in \refs{\nati-\ils} as well as some
generalizations.

In section 10 we summarize the results
and comment on them.
An appendix outlines the derivation
in gauge theory of one of the results
we deduce in the text from brane theory.

\bigskip
\newsec{The cast of characters}
\medskip

We will be mostly working in type IIA string
theory in ten dimensional Minkowski space.
The theory is non-chiral; denoting by $Q_L$,
$Q_R$ the space-time supercharges generated
by left and right moving worldsheet degrees
of freedom \ref\fms{D. Friedan, E. Martinec
and S. Shenker, \np{271}{1986}{93}.}, we have:
\eqn\tend{\eqalign{
\Gamma^0\cdots\Gamma^9 Q_L=& +Q_L\cr
\Gamma^0\cdots\Gamma^9 Q_R=& -Q_R\cr
}}
Type IIA string theory is actually an eleven
dimensional theory. The eleventh dimension,
$x^{10}$, is a circle of radius $R_{10}$ proportional
to the string coupling $\lambda$. In the
M-theory limit, $\lambda\to\infty$,
eleven dimensional Lorentz invariance is
restored, and the two spinors $Q_R$, $Q_L$
in \tend\ combine into a single spinor, the
${\bf 32}$ of $SO(10,1)$.

We will study three kinds of objects, each of
which separately preserves half of the SUSY
of the theory:

\item{(a)} Neveu-Schwarz (NS), or solitonic,
fivebranes, with tension proportional to
$1/\lambda^2$. These branes couple
magnetically to the NS $B_{\mu\nu}$ field
and are thus magnetic duals of fundamental
IIA strings. In M-theory the NS fivebrane
is interpreted as a ``transverse'' fivebrane,
living at a point in the eleventh dimension
$x^{10}$. At weak string coupling, the
configuration of an NS fivebrane in flat
Minkowski spacetime has been studied by
Callan, Harvey and Strominger \ref\chs{C. Callan,
J. Harvey and A. Strominger, \np{359}{1991}{611};
\np{367}{1991}{60}.} who constructed
the appropriate CFT for this situation.
This CFT is not well understood, especially
close to the core of the brane, where the
dilaton formally blows up and the CFT description
is not reliable, even when the string coupling far
from the fivebrane is weak. The worldvolume field
theory has (2,0) SUSY in six dimensions;
the massless spectrum includes \chs\ a self-dual
Kalb-Ramond field and five scalars parametrizing
fluctuations of the fivebrane in
the five transverse dimensions of M-theory.

\item{(b)} Dirichlet (D) $p$-branes, with tension
proportional to $1/\lambda$. In type IIA string
theory there are branes with $p=0,2,4,6,8$,
which couple to appropriate $p+1$ form gauge fields
in the Ramond sector. We will be mainly interested
in four and six branes, although other branes
will also play a role, providing non-perturbative
corrections to the classical dynamics. The D
fourbrane corresponds in M-theory to a
``longitudinal'' fivebrane
wrapped around $x^{10}$. The D sixbrane
can be thought of as a Kaluza-Klein monopole.
At weak string coupling, D branes are governed by
a CFT which has been constructed by Green
\ref\mbg{M. B. Green, \np{103}{1976}{333};
\pl{69}{1977}{89}; {\bf 201B} (1988) 42; 
{\bf 282B} (1992) 380; {\bf 329B} 1994 435.},
Polchinski
\ref\polch{J. Polchinski, hep-th/9510017,
\prl{75}{1995}{4724}.}, and others
\ref\revie{For a review and
further references to the original work, 
see \eg\ J. Polchinski, hep-th/9611050.},
and is rather well understood. The
worldvolume theory on an infinite Dirichlet
$p$-brane is a $p+1$ dimensional field
theory invariant under sixteen supercharges; 
the massless spectrum includes a $p+1$ 
dimensional $U(1)$ gauge field and $9-p$ 
scalars parametrizing fluctuations of the 
$p$-brane in the transverse dimensions. For
$N$ parallel $p$-branes, both the gauge field and
$9-p$ scalars are promoted to $N\times N$ matrices
\ref\eww{E. Witten, hep-th/9510135,
\np{460}{1996}{335}.} giving rise to $U(N)$ gauge theory
on the worldvolume with $9-p$ adjoint scalar fields,
and the fermions needed for supersymmetry.

\item{(c)} Orientifolds (O) are objects that at
least at weak string coupling are not dynamical
\revie; otherwise, they are similar to D branes. An
orientifold $p$-plane breaks the same half of
the SUSY as a parallel Dirichlet $p$-brane. 
It carries charge under the same Ramond
$p$ form gauge fields  as such a D brane. The
charge of an orientifold $p$-plane is equal
and opposite to that of $2^{p-5}$ {\it physical}
Dirichlet $p$-branes. In the presence of an
orientifold plane all D branes (which are 
outside the orientifold) acquire mirror
images; thus the number of such D branes 
is double the number of physical branes.

\noindent
Clearly, there are many ways to construct
configurations of branes preserving four
of the supercharges (1/8 of the original
supersymmetry). For our purposes it will
be sufficient to consider a class of
configurations built out of four kinds of
differently oriented  branes, and two kinds
of orientifold planes:

\item{(1)} NS fivebrane with worldvolume
$(x^0, x^1, x^2, x^3, x^4, x^5)$, which lives at
a point in the $(x^6, x^7, x^8, x^9)$ directions.
The NS fivebrane preserves supercharges of the
form
$\epsilon_LQ_L+\epsilon_RQ_R$, with
\eqn\nsfive{
\eqalign{\epsilon_L=&\Gamma^0\Gamma^1
\Gamma^2\Gamma^3\Gamma^4\Gamma^5\epsilon_L\cr
\epsilon_R=&\Gamma^0\Gamma^1\Gamma^2
\Gamma^3\Gamma^4\Gamma^5\epsilon_R.\cr
}}

\item{(2)} A differently oriented NS fivebrane,
which we will refer to as the \nsp\ fivebrane. Its
worldvolume is $(x^0, x^1, x^2, x^3, x^8, x^9)$; 
it lives at a point in the $(x^4, x^5, x^6, x^7)$ 
directions, and preserves the supercharges
\eqn\nsprime{
\eqalign{\epsilon_L=&
\Gamma^0\Gamma^1\Gamma^2\Gamma^3
\Gamma^8\Gamma^9\epsilon_L\cr
\epsilon_R=&\Gamma^0
\Gamma^1\Gamma^2\Gamma^3
\Gamma^8\Gamma^9\epsilon_R.\cr
}}

\item{(3)} D sixbrane with worldvolume
$(x^0, x^1, x^2, x^3, x^7, x^8, x^9)$,
which lives at a point in the $(x^4, x^5, x^6)$
directions. The D sixbrane preserves
supercharges satisfying
\eqn\dsix{
\epsilon_L=\Gamma^0\Gamma^1
\Gamma^2\Gamma^3\Gamma^7\Gamma^8
\Gamma^9\epsilon_R.
}

\item{(4)}
D fourbrane with worldvolume
$(x^0, x^1, x^2, x^3, x^6)$,
which lives at a point in the 
$(x^4, x^5, x^7, x^8, x^9)$ directions,
and preserves supercharges satisfying
\eqn\dfour{
\epsilon_L=\Gamma^0\Gamma^1
\Gamma^2\Gamma^3\Gamma^6\epsilon_R.
}

\item{(5)} Orientifold sixplane,
at a point in the $(x^4, x^5, x^6)$
directions. As discussed above, it
behaves similarly to the D sixbrane
of item (3).

\item{(6)} Orientifold fourplane,
at a point in the
$(x^4, x^5, x^7, x^8, x^9)$
directions, which behaves similarly
to the D fourbrane of item (4).

\noindent
It is easy to see that a configuration
including all the objects (1) -- (6)
preserves four supercharges. Indeed,
the second equations in \nsfive, \nsprime\
are satisfied by four of the sixteen
independent spinors $\epsilon_R$. $\epsilon_L$
is determined by (say) \dsix, and one can
check that the first equations in \nsfive,
\nsprime\ and \dfour\ are identities.

\nref\bdl{M. Berkooz, M. Douglas and R. Leigh,
hep-th/9606139, \np{480}{1996}{265}.}%
\nref\barbon{J. Barbon, hep-th/9703051.}%
It should be mentioned for completeness
that more general configurations preserving
$N=1$ supersymmetry can be constructed by
adding NS fivebranes, D sixbranes and O6 planes  
which are rotated with respect to the original 
ones in the $(x^4, x^5, x^8, x^9)$ directions 
(see \refs{\bdl, \barbon} for recent discussions). 
We will not discuss such configurations here, 
except for a few scattered comments.

\bigskip
\newsec{Brane configurations and some of
their classical properties}
\medskip

The field theory on an infinite
fourbrane is a five dimensional
gauge theory invariant under
sixteen supercharges. We will study,
following \hw, the theory on fourbranes
stretched between fivebranes. In this
case, the boundary conditions eliminate
some of the physical fluctuations of the
infinite fourbrane, and reduce the amount
of SUSY of the theory.

Consider, for example, a single
D fourbrane stretched along the
$x^6$ direction between two
NS fivebranes \hw. Since the ends of the
fourbrane are constrained
to lie on the fivebranes, the
modes corresponding to
fluctuations of the fourbrane in the
$(x^6, \cdots, x^9)$ directions
are massive. The configuration
preserves eight supercharges
(those satisfying \nsfive, \dfour),
and describes at long distances
a four dimensional $N=2$ supersymmetric
$U(1)$ gauge theory\foot{In a recent paper
\ref\ewew{E. Witten, hep-th/9703166.},
E. Witten pointed out that this $U(1)$
is in fact frozen (\ie\ its gauge coupling
vanishes) due to a divergence which has an
infrared interpretation on the fivebrane
and an ultraviolet one on the fourbrane.
This does not modify the present discussion
(or the constructions in the rest of the paper)
since we can cancel the divergence in question
by adding additional fourbranes attached to
the NS fivebranes ``at infinity'' in $(x^4, x^5)$.
Since all states charged under the $U(1)$ gauge
fields on both the original and the new fourbranes
are arbitrarily heavy, we can ignore them.}.
The position of the D fourbrane on the NS
fivebranes in the $(x^4, x^5)$ plane
corresponds to the expectation
value of the scalar in the $U(1)$ vector
multiplet. As is clear from
the brane description, physics
is independent of this
expectation value -- the scalar
does not carry $U(1)$ charge.
The distance between the
two NS fivebranes in the $x^6$
direction, $L_6$, determines the gauge
coupling of the $N=2$ supersymmetric gauge
theory on the brane, $1/g^2\propto
L_6/\lambda$.

Consider now adding a D sixbrane
at a point in the $x^6$ direction
that is between the locations
of the two NS fivebranes. The
additional brane does not break
any further SUSY \hw. The
fourbrane worldvolume dynamics
corresponds now to $N=2$ supersymmetric
$U(1)$ gauge theory with a
single charged hypermultiplet,
which arises from 4 -- 6 strings.
The relative position of the
D fourbrane and D sixbrane
in the $(x^4, x^5)$ directions
determines the mass of the
hypermultiplet.

The relative position of
the two NS branes in the $(x^7, x^8, x^9)$
directions plays the role of a
Fayet-Iliopoulos (FI) D-term in the $U(1)$
gauge theory. From the brane
configuration it is clear that only
when the two NS fivebranes are
at the same value of $(x^7, x^8, x^9)$,
can the D fourbrane stretch between them
without breaking SUSY. This is
a reflection of the fact that in the
presence of a non-vanishing FI D-term,
the gauge theory has no supersymmetric
vacuum with unbroken gauge symmetry.
For non zero
D-terms there is an isolated vacuum,
in which the theory is in a Higgs phase.
The interpretation of this in brane theory
is the following \hw: although
for non zero FI D-terms a D fourbrane
cannot connect two NS fivebranes,
it can break into two pieces, each of which
connects the D sixbrane to one of
the two NS fivebranes. Since
$(x^7, x^8, x^9)$ are part of the
worldvolume of the D sixbrane,
as the two NS branes are separated
in these directions, the two
pieces of the original fourbrane
could separate in the $(x^7, x^8, x^9)$
directions while staying on the D sixbrane.

Thus we learn that the Higgs mechanism
is described in brane theory
as the breaking of D fourbranes
stretched between NS branes, on
D sixbranes. The fact that in the $U(1)$
gauge theory considered above the
vacuum one finds for non zero D-terms is
isolated, implies that there are no massless
modes associated with fourbranes stretched
between D sixbranes and NS fivebranes. This
is consistent with the fact that such a D
fourbrane cannot fluctuate in the
$(x^4, \cdots, x^9)$ directions
due to the boundary conditions on the two
ends. $N=2$ SUSY implies that there is also
no massless worldvolume gauge field on it.

An interesting qualitatively new effect
arises when we generalize the previous
discussion to the case of $N_c>1$ fourbranes
stretched between two NS fivebranes
(still in the presence of one D sixbrane
between the NS branes). This configuration
describes an $N=2$ supersymmetric $U(N_c)$
gauge theory with a hypermultiplet in the
fundamental representation of the gauge group.
If we now displace one of the NS fivebranes
relative to the other in the $(x^7, x^8, x^9)$
directions, turning on a FI D-term, naively the
previous discussion applies, and the $N_c$
fourbranes can break on the single D sixbrane
to yield an isolated supersymmetric vacuum for
all $N_c$. A gauge theory analysis reveals that
this conclusion is false -- for all $N_c>1$ the
F and D term conditions for a supersymmetric
vacuum have no solution. This and many related
puzzles would be resolved if the following
``s-rule'' held \hw:

\item{}{\it A configuration in which an NS
fivebrane and a D sixbrane are connected by
more than one D fourbrane is not supersymmetric.}

\noindent
A geometric explanation of the s-rule has been
proposed in \ov. In brane dynamics it may be
related to the fact that two fourbranes connecting
a given NS fivebrane to a given D sixbrane are
necessarily on top of each other -- a rather
singular situation.

Later in the paper we will be interested
in moving NS fivebranes through D sixbranes.
At first sight this appears to be a
singular procedure, since when the two kinds
of branes are at the same value of $x^6$,
they actually meet in spacetime. In \hw\
it was proposed that when an NS fivebrane
passes through a D sixbrane in the $x^6$
direction, a third brane is created, a D
fourbrane connecting the two original
branes. It was pointed out that this
creation of a brane is necessary to preserve
magnetic charge. More importantly for us,
in \hw\ evidence was presented that this 
process also implies that the low energy 
dynamics is insensitive to the relative 
position of D sixbranes and the NS fivebrane. 
Thus, despite appearances, the process of moving 
an NS fivebrane through a D sixbranes is
smooth. This is going to play a role in the
construction of the next sections.

In the next sections we will be interested
in configurations which include \nsp\ 
fivebranes; hence, it is useful to repeat the 
above discussion with \nsp\ fivebranes replacing 
NS fivebranes. We will next mention a few of the 
main points.

Consider a configuration of $N_c$ fourbranes 
stretched between two \nsp\ fivebranes along 
the $x^6$ direction, with $N_f$ sixbranes 
located between the two \nsp\ fivebranes.
Unlike the previous case, this configuration 
preserves four supercharges, and describes at 
long distances a four dimensional $N=1$ 
supersymmetric $U(N_c)$ gauge theory with one 
adjoint chiral multiplet, $X$, and $N_f$ 
fundamental multiplets, $Q, \tilde Q$ (we 
suppress flavor indices). The difference with 
the previous case is that the superpotential
$W=\tilde Q X Q$ required by $N=2$ SUSY in $4d$ 
is absent here. The resulting model is in fact 
one of the simplest examples of $N=1$ supersymmetric 
gauge theory where the theory is known to be in 
a non abelian Coulomb phase for all $N_f\ge1$, but 
the long distance dynamics is not understood (see e.g. 
the third reference in \ks\ for some comments on this 
model).

Fluctuations of the $N_c$ fourbranes inside the
\nsp\ fivebrane in the $(x^8, x^9)$ plane
parametrize the Coulomb branch of the model.
Displacements of the $N_f$ D sixbranes relative
to the \nsp\ fivebranes in the $(x^4, x^5)$
directions give masses to the fundamental multiplets
$Q$, $\tilde Q$. The relative position of the two
\nsp\ fivebranes in the $(x^4, x^5)$ directions is
not an independent parameter, as it can be compensated
by changing the positions of the D sixbranes in the
$(x^4, x^5)$ plane (\ie\ the masses of the fundamentals),
and an overall rotation of the configuration.

The relative displacement of the two \nsp\
fivebranes in the $x^7$ direction plays the
role of a FI D-term; note that unlike the
$N=2$ SUSY case discussed before, here the
D-term is a single real number, in agreement
with field theory expectations. Separating the
two \nsp\ fivebranes in the $x^7$ direction
(\ie\ turning on a FI D-term) should not break
SUSY for all $N_c$, $N_f>0$. In fact, a gauge
theory analysis leads to the conclusion that
complete Higgsing is possible for all $N_f\ge1$;
the (complex) dimension of the Higgs branch is:
$2N_fN_c+N_c^2-N_c^2=2N_fN_c$. The first two terms
on the left hand side are the numbers of components
in the fundamental and adjoint chiral multiplets,
and the negative term accounts for degrees of
freedom eaten up by the Higgs mechanism.

The brane configuration provides
a simple picture of the moduli
space of vacua. As we have learned
before, complete Higgsing
corresponds to breaking all $N_c$
fourbranes on various D sixbranes.
The resulting fourbranes can then
further break on other D sixbranes;
a generic point in the Higgs branch
corresponds to splitting each of the $N_c$
fourbranes into $N_f+1$ pieces,
the first connecting the left
\nsp\ fivebrane to the first D
sixbrane, the second connecting the first
two D sixbranes etc. To calculate
the dimension of the moduli space
one notes that:

\item{(a)} A fourbrane stretched between two
D sixbranes has two complex
massless degrees of freedom,
corresponding to fluctuations
of the brane in the $(x^7, x^8, x^9)$
directions, and $A_6$, the compact
component of the fourbrane
worldvolume  gauge field.

\item{(b)} A fourbrane stretched between an 
\nsp\ fivebrane and a D sixbrane has one 
complex massless degree of freedom, corresponding
to fluctuations of the brane in the $(x^8, x^9)$ 
directions.

\item{(c)} One does not expect an analog of 
the s-rule for fourbranes stretched between an 
\nsp\ fivebrane and a D sixbrane, \eg\ because
two such fourbranes can be separated in the 
$(x^8, x^9)$ directions, which are common to 
both kinds of branes.

\noindent
Using these rules we find that the dimension of 
moduli space of brane configurations with 
completely broken $U(N_c)$ gauge symmetry is: 
$2N_c(N_f-1)+2N_c=2N_cN_f$, in agreement with the
gauge theory analysis. Of course, one can read the 
above arguments in the opposite direction; the 
agreement of the gauge theory analysis with the brane 
configuration counting is strong evidence for the 
validity of assumptions (a) -- (c) about brane dynamics.

\bigskip
\newsec{Some quantum properties of branes}
\medskip

One of the interesting features of brane
configurations preserving $N=1$ SUSY in
$4d$ (or $N=2$ in $3d$) is the fact that
the classical picture of branes connected
at right angles and not exerting any forces
on each other is corrected quantum
mechanically\foot{In contrast, in theories
with eight or more supercharges, branes do
not exert forces on each other quantum
mechanically as well, since there are no
superpotentials. There are of course
other interesting quantum effects \hw, 
\ewew.}. In this section we will describe the 
quantum bending of branes \ewew, and formulate 
some phenomenological rules regarding the quantum
forces between different fourbranes. These
forces will be deduced by comparison to gauge
theory. One of the main advantages of the brane
description is its universality; interactions
deduced in a particular situation can be used
in others to learn about strongly coupled gauge
dynamics.

\medskip
\subsec{Bending of NS fivebranes}

Consider a configuration where a
D fourbrane ends on
an NS fivebrane from the left\foot{Here and
below, positions of various branes
refer to the $x^6$ direction, unless
stated otherwise.}.
The fourbrane is actually an M-theory
fivebrane spanning the complex
$s=x^6+ix^{10}$ plane (in addition
to $(x^0, x^1, x^2, x^3)$ which we
will suppress here), and is located
at fixed $v=x^4+ix^5$, say $v=a$. The NS
fivebrane fills the $v$ plane, and is
classically located at a fixed value of
$s$.

Quantum mechanically \ewew, the NS fivebrane
bends away from the D fourbrane. Far from the
point at which classically the fourbrane meets 
the fivebrane, $v=a$, the location of the latter 
is described by:
\eqn\sfive{s_5=R_{10}\ln(v-a)}
This bending can be thought of as a quantum
effect in the type IIA string description.
Equation \sfive\ implies that the classical symmetry
of continuous rotation in the $v$ plane,
$(v-a)\to e^{i\alpha}(v-a)$, is broken quantum
mechanically. This is the geometrical M-theory 
realization of the gauge theory anomaly in a 
chiral global $U(1)$ current.

\medskip
\subsec{Three dimensional N=2 supersymmetric
gauge theory}

\nref\bhoy{J. de Boer, K. Hori,
Y. Oz and Z. Yin, hep-th/9702154.}%
\nref\bho{J. de Boer, K. Hori and Y. Oz,
hep-th/9703100.}%
\nref\ahiss{O. Aharony, A. Hanany,
K. Intriligator, N. Seiberg and M.
Strassler, hep-th/9703110.}%
\noindent
To deduce the quantum forces between branes
we will compare some recent discussions of
three dimensional $N=2$ supersymmetric
gauge theories \refs{\bhoy-\ahiss}
with the brane picture. We start by describing
the relevant gauge theory results.

Consider a four dimensional $N=1$
supersymmetric $U(N_c)$ gauge theory
coupled to $N_f=N_c$ flavors of quarks
$Q^i$, $\tilde Q_{\tilde i}$ $(i, \tilde
i=1,\cdots, N_f)$, in the fundamental
representation of $U(N_c)$ (the
``electric'' theory). Compactify
one of the dimensions on a circle of radius
$R$; this gives rise at energies much lower
than $1/R$ to a three dimensional $N=2$
supersymmetric gauge theory. Classically,
this model has an $N_c(=N_f)$ 
dimensional\foot{Here and below we always 
count complex dimensions of moduli spaces.}
Coulomb branch and a $2N_cN_f-N_c^2=N_c^2$
dimensional Higgs branch which meet at the origin,
as well as mixed Higgs -- Coulomb branches
in which part of the gauge symmetry is
unbroken. One can also add ``real masses''
to the quarks which split the singularities
in which the Coulomb and Higgs branches meet.
We will discuss their role in the brane
picture later.

Quantum mechanically, the low energy behavior
of this theory (in the absence of real masses)
is reproduced \ahiss\ by a sigma model
for $N_c^2+2$ chiral superfields $V_\pm$, 
$M^i_{\tilde i}$, with the superpotential
\eqn\qsp{W=V_+V_-(\det M-\eta)}
where $\eta$ is related to the three and
four dimensional gauge couplings:
$\eta\sim e^{-1/Rg_3^2}\sim e^{-1/g_4^2}$.
In the four dimensional limit $R\to\infty$,
$\eta$ turns into the four dimensional
QCD scale $\eta\to\Lambda_4^{2N_c}$, while
in the three dimensional limit $R\to0$
(with $g_3$ fixed), $\eta\to 0$.
$M$ should be thought of as representing
the meson field $M^i_{\tilde i}=Q^i\tilde
Q_{\tilde i}$ and $V_\pm$ parametrize the
Coulomb branch.

Varying \qsp\ w.r.t. the fields $V_\pm$, $M$ 
gives rise to the equations of motion
\eqn\eqq{V_\pm(\det M-\eta)=0; \;\; V_+V_-
(\det M) (M^{-1})_i^{\tilde i}=0}
Consider first the three dimensional case
($R=\eta=0$). There are three branches of
moduli space:
\noindent
\item{1)} $V_+=V_-=0$; $M$ arbitrary.
\noindent
\item{2)} $V_+V_-=0$; $M$ has rank
at most $N_c-1$.
\noindent
\item{3)} $V_+$, $V_-$ arbitrary; $M$ has
rank at most $N_c-2$.

\noindent
The first branch can be thought of as a Higgs
branch, while the last two are mixed Higgs -- Coulomb
branches. The three branches meet on a complex
hyperplane on which the rank of $M$ is $N_c-2$
and $V_+=V_-=0$. Most of the classical $N_c$
dimensional Coulomb branch is lifted; its
only remnants are $V_\pm$.

The understanding of the theory with $N_f=N_c$
allows us to study models with any $N_f\le N_c$
by adding masses to some of the flavors and 
integrating them out.
Adding a complex quark mass term $W=-m M $ to \qsp,
the following structure emerges. If the rank of
$m$ is one, one finds after integrating out the
massive flavor a moduli space of vacua with
$V_+V_-\det M=1$ ($M$ is the $(N_f-1)\times
(N_f-1)$ matrix of classically massless mesons).
If the rank of $m$ is larger than one, one finds
a superpotential with a runaway behavior.
For example, if we add two non-vanishing masses:
\eqn\wmas{W=V_+V_-\det M-m_1^1M^1_1-m_2^2M^2_2}
we find, after integrating out the massive mesons
$M_1^i$, $M_j^1$, $M_2^i$, $M_j^2$:
\eqn\wintout{W=-{m_1^1m_2^2\over V_+V_-\det M}}
where, again, $M$ represents the $(N_f-2)^2$
classically massless mesons. Clearly, the
superpotential \wintout\ does not have a minimum
at finite values of the fields; there is no stable 
vacuum.

When the radius of the circle is not strictly zero
($\eta\not=0$), the analysis of \eqq\ changes somewhat.
There are now only two branches:
\noindent
\item{1)} $V_+=V_-=0$; $M$ arbitrary.
\noindent
\item{2)} $V_+V_-=0$; $\det M=\eta$.

\noindent
In particular, there is no analog of the
third branch of the three dimensional
problem. The two branches meet on a complex
hyperplane on which $\det M=\eta$ and
$V_+=V_-=0$. The structure for all
$\eta\not=0$ agrees with the four dimensional
analysis \refs{\neqone, \nonerev}.

If we add to \qsp\ a complex mass term
$W=-mM$ with a mass matrix $m$ whose rank is
smaller than $N_f$, the vacuum is destabilized
(including the case of a mass matrix of rank
one where previously there was a stable vacuum).
If the rank of $m$ is $N_f$, so that
the low energy theory is pure $U(N_c)$
SYM, there are
$N_c(=N_f)$ isolated vacua which run off to
infinity as the radius of the circle $R$ goes
to zero (there is also a decoupled moduli space
for the $U(1)$ piece of the gauge group).

Another theory which will prove useful
is the compactified $N=1$ SYM theory
with gauge group $U(N_c)$ and $N_f=N_c$
flavors of quarks described above, coupled
to a gauge singlet superfield $N_i^{\tilde i}$
via a Yukawa interaction:
\eqn\wmn{W=N_i^{\tilde i} Q^i\tilde Q_{\tilde i}
=N_i^{\tilde i}M^i_{\tilde i}}
Classically, this ``magnetic'' theory has
an $N_c^2+N_c$ dimensional moduli space of vacua,
corresponding to giving the singlet meson $N$
and the adjoint scalar (together with the dual
of the gauge field) arbitrary expectation values,
keeping $Q=\tilde Q=0$. At generic points on this
moduli space the gauge group is broken to
$U(1)^{N_c}$.

To study the quantum mechanical situation, we
add the classical superpotential \wmn\ to
\qsp. Varying the resulting superpotential
with respect to the fields $M$, $N$, $V_\pm$
gives rise to the equations:
\eqn\eqqn{V_\pm\eta=0;\;\;M=0;\;\;N=0}
Thus, in the three dimensional case $\eta=0$
we find a two complex dimensional Coulomb
phase parametrized by $V_+$, $V_-$, while for
non-zero $R$ there is an isolated vacuum at
the origin. The classical $N_c^2$ dimensional
moduli space parametrized by $N$ is completely
lifted. For non-zero $R$ the Coulomb branch is
lifted as well, while for $R=0$ a two dimensional
subspace remains.

To understand the gauge theory results described
above in brane theory, we next turn to the
construction of these gauge theories using branes.

\medskip
\subsec{The brane construction of classical 3d N=2 SQCD}

Consider a configuration of $N_c$ D fourbranes
stretched between an NS fivebrane and an \nsp\
fivebrane along the $x^6$ direction, with the
geometry described in section 2. This configuration
preserves four supercharges, and corresponds
at distances much larger than the separation
between the NS and \nsp\ branes, $L_6$, to a
four dimensional $N=1$ supersymmetric theory on
the worldvolume of the fourbranes.

In analogy to section 3, it is clear that the
boundary conditions on the two ends of each
fourbrane suppress long wavelength fluctuations
of the fourbranes in the $(x^4, \cdots, x^9)$
directions. The only massless worldvolume degrees
of freedom are the $U(N_c)$ gauge bosons and their
superpartners. To add matter, we insert $N_f$ D
sixbranes at values of $x^6$ that are between those
corresponding to the positions of the NS and \nsp\
branes. 4 -- 6 strings describe $N_f$ chiral
multiplets in the fundamental representation of
$U(N_c)$, $Q^i, \tilde Q_i$.

To compare to the electric gauge theory described in
the previous subsection we compactify one of the
fourbrane worldvolume dimensions, say $x^3$, on a
circle of radius $R$. It is convenient to  perform
a T-duality transformation which transforms type IIA
to type IIB and turns the D fourbrane wrapped around
$x^3$ into a D threebrane at a point on the dual
circle of radius $R_3=1/R$. The NS and \nsp\
fivebranes transform to themselves under this
T-duality, while the D sixbrane turns into a D 
fivebrane at a point in $(x^3, x^4, x^5, x^6)$. 
The positions of the different D fivebranes in $x^3$ 
play the role of the real masses of the quarks. 
As $R_3\to\infty$ we recover the three dimensional 
$N=2$ SQCD theory with $\eta=0$ discussed in the 
previous subsection, while the four dimensional 
limit corresponds to $R_3\to 0$.

The magnetic theory with the singlet meson $N$ \wmn\
is obtained from a different but related
configuration. We place an \nsp\ fivebrane
to the left of an NS fivebrane and add $N_f$
D fivebranes to the left of the \nsp\ fivebrane.
Each of the D fivebranes is connected to the
\nsp\ fivebrane by a D threebrane, and there are $N_c$
threebranes stretched between the \nsp\ and NS 
fivebranes\foot{We will discuss the four dimensional 
version of this theory in section 5.}. The
quarks arise as 3 -- 3 strings stretched
between the two kinds of threebranes, while 
the meson $N$ arises from 3 -- 3 strings 
connecting threebranes stretched between 
the \nsp\ fivebrane and different D 
fivebranes. The classical moduli space 
obtained by giving independent expectation 
values to all components of $N$ is realized
in the brane picture by breaking all $N_f$ 
D threebranes connecting the \nsp\ fivebrane 
to the D fivebranes in all possible ways.

\medskip
\subsec{Three dimensional gauge theory and brane
interactions}

The quantum gauge theory effects described above
are due in the brane picture to two basic quantum
properties of threebranes:

\noindent
\item{(1)} There is a long range interaction
between a D threebrane stretched between an
\nsp\ fivebrane and an NS fivebrane, and any
other threebrane ending on the same \nsp\
fivebrane. The interaction is repulsive if
the two threebranes end on the same side of
the \nsp\ fivebrane; it is attractive if one
ends on the \nsp\ fivebrane from the left
and the other from the right.

\noindent
\item{(2)} The repulsion between different
threebranes ending on the same side of an \nsp\
fivebrane can be ``screened'' by D fivebranes
located between the threebranes (in $x^3$). 
This screening is effective if and only if 
the threebrane stretched between NS and \nsp\ 
fivebranes mentioned in (1) can break on the D 
fivebrane before reaching the other threebrane.

\noindent
Clearly, by T-duality there are similar interactions
between fourbranes ending on an \nsp\ fivebrane
in type IIA string theory. In the rest of this
subsection we will explain how the two rules stated
above encode different gauge theory phenomena described
in section {\it 4.2}. Later we will apply these rules
to more general situations in three and four dimensions
and will see that they pass some additional consistency
tests.

Consider first the electric theory of section {\it 4.2}.
If the mass matrix $m$ has rank $N_c$, the theory
describes $N=1$ supersymmetric $U(N_c)$ gauge theory
with no massless matter, on $R^3\times S^1$.
As we saw, if the radius of the circle, $R$, vanishes, the
theory has no stable vacuum due to a runaway superpotential
on the classical Coulomb branch. When $R>0$ the classical
moduli space is replaced by $N_c$ isolated vacua which run
off to infinity as $R\to0$.

In the brane picture these results are consequences of
the repulsion between threebranes stretched between
NS and \nsp\ fivebranes. Recall that the threebranes
live on a dual circle with radius $R_3=1/R$. For
finite $R_3$ the branes arrange around the circle
at equal spacings, maximizing the distances
between them and leading to an isolated vacuum.
It is clear that the $N_c$ dimensional
Coulomb phase is lifted, except for the decoupled
$U(1)$ Coulomb branch which appears as the freedom
to rigidly rotate all $N_c$ threebranes around the
$x^3$ circle. The fact that there are $N_c$ vacua 
has to do with the dual of the three dimensional
gauge field, and is not expected to be seen
geometrically (in $3d$). It is also clear that in the
three dimensional limit $R_3\to\infty$ the vacua run
off to infinity, due to the long range repulsion.

The repulsive potential between a pair of threebranes 
can be thought of in this case\foot{Note, however, that
this explanation does not apply to some of the other
attractive and repulsive interactions between threebranes
implied by the rules above.} as due to Euclidean 
D strings stretched between the NS and \nsp\ fivebranes 
and between the two D threebranes \hw, \bhoy\ (which 
are instantons in the low energy three dimensional 
gauge theory). Since there are two fermionic zero 
modes in the presence of these instantons, they lead 
to a superpotential on the classical Coulomb branch. 
For D threebranes stretched between NS fivebranes  
(discussed in \hw) the same instantons contribute to 
the metric on the Coulomb branch, since there are then 
four zero modes in the presence of the instanton.

In situations when the rank of the mass matrix is
smaller than $N_c$, there are massless quarks in the
gauge theory; in the brane description, there are
D fivebranes that can ``screen'' the interactions
between the threebranes. This screening can be seen
directly by studying the Euclidean D strings stretched
between D threebranes. If the worldsheet of such a D
string intersects a D fivebrane, two additional zero
modes appear and the contribution to the superpotential
vanishes.

Consider for example the case where the number 
of massless flavors is $N_f=N_c-2$ in the three
dimensional limit $R=0$. We saw in section
{\it 4.2} that this theory develops a runaway 
quantum superpotential \wintout. In the brane 
picture, we have $N_c$ threebranes stretched 
between NS and \nsp\ fivebranes, and $N_c-2$ D 
fivebranes located at the same value of $x^3$ 
(we are restricting to the case of vanishing 
real masses for now) between the NS and \nsp\ 
fivebranes.

Due to the repulsion between unscreened threebranes
stretched between NS and \nsp\ fivebranes, $N_c-2$
of the $N_c$ threebranes must break on different
D fivebranes. The s-rule implies that once this has
occurred, no additional threebranes attached to the
NS fivebrane can break on these D fivebranes.
We are left with two unbroken threebranes, one
on each side of the D fivebranes (in $x^3$).
These threebranes repel each other, as well as
the pieces of the broken threebranes stretched
between the \nsp\ fivebrane and the D fivebrane
closest to it. There is no screening in this
situation since all $N_c-2$ D fivebranes are
connected to the NS fivebrane; hence the two threebranes
stretched between the NS and \nsp\ fivebranes cannot
break on these D fivebranes. The system is unstable,
and some or all of the threebranes mentioned above
must run away to infinity.

This is in agreement with the gauge theory analysis of the
superpotential \wintout. One can think of $V_\pm$ as the
positions in $x^3$ of the two threebranes stretched
between NS and \nsp\ fivebranes mentioned above (as usual,
together with the dual of the three dimensional gauge field).
The potential obtained from  \wintout\ indeed suggests a
repulsion between the different threebranes.

It is clear that the arguments above continue to hold
when the radius of the circle on which the threebranes
live is finite. While the two threebranes stretched between
the NS and \nsp\ fivebranes can no longer run away to
infinity in the $x^3$ direction, those connecting the
\nsp\ fivebrane to a D fivebrane (representing components
of $M$) can, and there is still no stable vacuum.
This is in agreement with gauge theory; adding the term
$W=\eta V_+V_-$ to \wintout\ and integrating out $V_\pm$
leads to a superpotential of the form $W\sim (\det M)^{-1/2}$.

The above discussion can be repeated with the same
conclusions for all values $1\le N_f\le N_c-2$.

For $N_f=N_c-1$ the gauge theory answer is different;
there is still no vacuum in the four dimensional case
$\eta\not=0$, while in three dimensions there is a
quantum moduli space of vacua with $V_+V_-\det M=1$.
In brane theory there are now $N_c-1$ D fivebranes,
and the interaction
between the D threebranes stretched between NS and \nsp\
fivebranes can be screened. Indeed, consider a situation
where $N_c-2$ of the $N_c$ threebranes stretched between
NS and \nsp\ fivebranes break on D fivebranes.
This leaves two threebranes and one D fivebrane that is
not connected to the NS fivebrane. If $R_3=\infty$
(\ie\ $\eta=0$), the single D fivebrane
can screen the repulsion between the two threebranes.
If the threebrane is at $x^3=0$, then using the rules
in the beginning of this subsection we deduce that
any configuration where one of the threebranes is at
$x^3>0$ while the other is at $x^3<0$ is stable.
The locations in $x^3$ of the two threebranes give
the two moduli $V_\pm$.
Thus, the brane picture predicts correctly the existence
of the quantum moduli space and its dimension. The
precise shape of the moduli space (the relation
between $\det M$ and $V_+V_-$) is more difficult
to see geometrically; nevertheless, it is clear
that due to the repulsion there is no vacuum when
either $V_+$ or $V_-$ vanish.

If the radius of the fourth dimension $R$ is not zero,
the brane picture predicts a qualitative change in the
physics. Since $R_3$ is now finite, the two threebranes
stretched between NS and \nsp\ fivebranes are no longer
screened by the D fivebrane -- they interact through the
other side of the circle. Thus one of them has to
break on the remaining D fivebrane, and one remains
unbroken because of the s-rule. The repulsion between
that threebrane and the threebranes stretched between
the \nsp\ fivebrane and a D fivebrane which is no longer
screened leads to vacuum destabilization, in agreement
with the gauge theory analysis.

For $N_f=N_c$ (and vanishing real masses) the 
brane theory analysis is similar to
the previous cases, and the conclusions are again in
agreement with gauge theory. For $R_3=\infty$ one finds
three phases corresponding to a pure Higgs phase in which
there are no threebranes stretched between NS and \nsp\
fivebranes, and two mixed Higgs -- Coulomb phases in which
there are one or two threebranes stretched between the NS
and \nsp\ fivebranes; the locations of the threebranes in
$x^3$ are parametrized by $V_\pm$. When there are two
unbroken threebranes, they must be separated in $x^3$ by
the D fivebranes, which provide the necessary screening.

For finite $R_3$, the structure is similar, except
for the absence of the branch with two unbroken
threebranes, which is lifted by the same mechanism
to that described in the case $N_f=N_c-1$ above.

\medskip
\subsec{The magnetic brane configuration}

In section {\it 4.3} we have presented a brane
configuration realizing a $U(N_c)$ gauge theory
with $N_f$ flavors of quarks coupled to a singlet
meson $N^i_{\tilde i}$ via the superpotential
\wmn. The quantum moduli space of this gauge
theory with $N_f=N_c$ was analyzed at the end
of section {\it 4.2}. In this subsection we will
revisit this theory from the point of view of brane
dynamics.

The brane configuration describing this theory
contains $N_c$ threebranes connecting the \nsp\
to an NS fivebrane on its right (we will refer
to these as threebranes of the first kind),
and $N_c$ threebranes (which we will refer
to as threebranes of the second kind) connecting
it to $N_c$  D fivebranes on its left.

Classically, this configuration has an $N_c^2+N_c$
dimensional moduli space corresponding to
arbitrary breaking of the threebranes of the second
kind on D fivebranes, and to independent motions of
the threebranes of the first kind in the $x^3$
direction. The rules of section {\it 4.4} imply that
quantum mechanically threebranes of the first kind
have repulsive interactions among themselves, and
attractive interactions with threebranes of the
second kind. These together with the s-rule imply
that the theory has a unique vacuum at the origin
where all $N_c$ D threebrane are aligned and can be
thought of as stretching between the NS fivebrane
and the $N_c$ D fivebranes.

Looking back at equation \eqqn\ we discover a puzzle.
For $\eta\not=0$ (\ie\ finite $R_3$) the gauge theory
analysis is in agreement with the brane picture, yielding
a unique vacuum at the origin of moduli space
$V_\pm=M=N=0$. For $R_3=\infty$ (the three dimensional
limit) the gauge theory analysis yields a two dimensional
moduli space parametrized by $V_\pm$. If true, it would
imply that two of the threebranes of the first kind can
remain unattached to threebranes of the second kind,
without feeling any attraction to them or repulsion
from each other.
This is puzzling since the relevant physics in this
problem is the interplay between the attraction between
branes of different kinds and repulsion of branes of the
same kind, and the existence of such an unlifted branch
of moduli space should not be sensitive to whether
$R_3$ is finite or infinite.

The above puzzle leads us to propose that in fact the
magnetic brane configuration described above gives rise
quantum mechanically to the following sigma model:
\eqn\qmagth{W=V_+V_-(\det M-\eta) +N M +
V_+W_-+ V_- W_+}
This is the same as the quantum superpotential of the
gauge theory  \wmn\ except for the fact that the
fields $V_\pm$ are now massive even for $\eta=0$.
To give them a mass we had to introduce dynamical 
fields $W_\pm$, whose geometrical role in the brane
configuration is not clear. Of course, the description
\qmagth\ is a quantum one. Classically, the
fields $V_\pm $ are not well defined; as described
in detail above, they parametrize the potentially
massless modes on the Coulomb branch after most
of it has been lifted by quantum effects.

\bigskip
\newsec{Classical electric and magnetic N=1 SQCD}
\medskip

After the detour to three dimensions, we return in
this section to four dimensional $N=1$ supersymmetric
gauge theory. The main purpose of this and the next
section is to derive Seiberg's duality \nati\ in
$U(N_c)$ gauge theory with matter in the fundamental
representation, by embedding the problem in string
theory. To achieve that we will in
sections {\it 5.1, 5.2} study brane configurations
realizing the electric and magnetic theories;
in section {\it 5.3} we will show that,
in the presence of a small FI D-term which
breaks the gauge symmetry completely, the moduli spaces
of the two theories in fact provide two different
descriptions of a single smooth space -- the moduli
space of vacua of the underlying brane theory.
This explains the equivalence between the two theories
in the Higgs phase. In section 6 we will complete the
demonstration of duality by studying the
quantum effects, which become large when we turn off
the D-term and approach the origin of moduli space.

\medskip
\subsec{The electric theory}

Consider the configuration of $N_c$ D fourbranes
stretched between an NS fivebrane and an \nsp\ fivebrane
along the $x^6$ direction, with $N_f$ D sixbranes at
values of $x^6$ that are between those corresponding
to the positions of the NS and \nsp\ branes discussed
in section 3. It is instructive to relate the
supersymmetric deformations of the gauge theory to
parameters defining the brane configuration, using
the dictionary established in the previous sections:

\noindent
{\sl 1) Higgs moduli space:}

\noindent
In the gauge theory, the structure
of moduli space is as follows. For $N_f<N_c$, the
$U(N_c)$ gauge symmetry can be maximally broken to
$U(N_c-N_f)$. The complex dimension of the
moduli space of vacua is:
$2N_cN_f-\left( N_c^2-(N_c-N_f)^2\right)=N_f^2$.
For $N_f\ge N_c$ the gauge symmetry can be completely
broken, and the complex dimension of the moduli space
is $2N_cN_f-N_c^2$.

In the brane description, Higgsing corresponds
to splitting fourbranes on sixbranes \hw.
Consider, \eg, the case $N_f\ge N_c$ (the case
$N_f<N_c$ is similar). A generic point in moduli
space is described as follows. The first D fourbrane
is broken into $N_f+1$ segments connecting the
NS fivebrane to the first (\ie\ leftmost)
D sixbrane, the first D sixbrane to the second,
etc., with the last segment connecting the rightmost
D sixbrane to the \nsp\ fivebrane. The second D fourbrane
can now only be broken into $N_f$ segments, because of the
s-rule: the first segment must stretch between the NS
fivebrane and the {\it second} D sixbrane, with the
rest of the breaking pattern as before.

Recalling (see section 3) that a D fourbrane stretched
between two D sixbranes has two complex massless
degrees of freedom,
while a D fourbrane stretched between a D sixbrane and an
\nsp\ fivebrane has one complex
massless d.o.f., we conclude that the
dimension of moduli space is:
\eqn\higgsm{\sum_{l=1}^{N_c}\left[2(N_f-l)+1\right]
=2N_fN_c-N_c^2}
in agreement with the gauge theory result.

\noindent
{\sl 2) Mass deformations:}

\noindent
In gauge theory we can turn on a mass matrix
for the (s)quarks, by adding a superpotential
\eqn\wmq{W=m_i^jQ^i\tilde Q_j}
with $m$ an arbitrary $N_f\times N_f$ matrix of complex
numbers. In the brane configuration, masses correspond
to relative displacement of the D sixbranes and the
D fourbranes (or equivalently the \nsp\ fivebrane)
in the $(x^4, x^5)$ directions. The configuration
can be thought of as realizing
a superpotential of the form \wmq, with the mass matrix
$m$ satisfying the constraint
\eqn\mmdag{[m, m^\dagger]=0}
Thus, we can diagonalize $m, m^\dagger$ simultaneously;
the locations of the D sixbranes are the eigenvalues
of $m$.

Hence, the brane configuration describes only a
subset of the possible deformations of the gauge
theory. This is a rather common situation in string
theory, and we will encounter additional examples
of this phenomenon below. Note also that the condition
\mmdag\ appears as a consistency condition in $N=2$
supersymmetric gauge theories. Our theory is clearly
not N=2 supersymmetric; nevertheless,  it is not surprising
that the condition \mmdag\ arises, since one can think
of $m$ as the expectation value of a superfield in the
adjoint of the $U(N_f)$ gauge group on the D sixbranes.
The theory on the infinite sixbranes is invariant under
sixteen supercharges in the bulk of the worldvolume, and
while it is broken by the presence of the other
branes, it inherits \mmdag\ from the theory with more
supersymmetry.

\noindent
{\sl 3) FI D-term:}

\noindent
In the gauge theory it is possible to turn on a D-term for
$U(1)\subset U(N_c)$. For $N_f<N_c$ this breaks SUSY. For
$N_f\ge N_c$ there are supersymmetric vacua in which the gauge
symmetry is broken, and the system is in a Higgs phase.
In the brane description, the role of the FI D-term is played
by the relative displacement of the NS and \nsp\ fivebranes
in the $x^7$ direction. Clearly, when the two are at different
values of $x^7$, a fourbrane stretched between them breaks SUSY.
To preserve SUSY, all such fourbranes must break on D sixbranes,
which as we saw above is only possible for $N_f\ge N_c$ because
of the s-rule. Once all fourbranes have been split, there is no
obstruction to moving the NS and \nsp\ fivebranes to different
locations in $x^7$. Note that the Higgs phase depends smoothly
on the D-term. In the brane construction the reason is that once
all $N_c$ D fourbranes have been broken on D sixbranes in a
generic way, nothing special happens when the relative
displacement of the two fivebranes in $x^7$ vanishes.

\noindent
{\sl 4) Global symmetries:}

\noindent
Supersymmetric QCD with gauge group $SU(N_c)$ and $N_f$
quarks has the global symmetry
$SU(N_f)\times SU(N_f)\times U(1)_B\times U(1)_x\times U(1)_a$.
The two $SU(N_f)$ factors rotate the quarks $Q^i$,
$\tilde Q_i$, respectively; $U(1)_B$ is a vectorlike
symmetry, which assigns charge $+1$
($-1$) to $Q$ ($\tilde Q$).
$U(1)_x$ and $U(1)_a$ are $R$ symmetries
under which the gaugino is assigned
charge one, and the quarks $Q$,
$\tilde Q$ have charge 0 or 1.
Only one combination of the two $R$ symmetries
is anomaly free, but this quantum effect is not expected
to be visible in the classical brane construction (we
will derive it from brane dynamics in section 6) .

In our case, the baryon number symmetry is gauged.
The brane configuration has a manifest $SU(N_f)$
symmetry, which is a gauge symmetry on the D
sixbranes and a global symmetry on the D fourbranes.
The other $SU(N_f)$ is presumably
broken at short distances and emerges at long
distances as an accidental symmetry.
The symmetries $U(1)_x$, $U(1)_a$ are exact
symmetries in the brane picture. They correspond
to rotations in the $(x^4, x^5)$ and $(x^8, x^9)$
planes, $U(1)_{45}$, $U(1)_{89}$. From the discussion
of the mass deformations and Higgs moduli space above,
it is clear that the mass parameters \wmq\ are
charged under $U(1)_{45}$, while the quarks
$Q$, $\tilde Q$ are charged under $U(1)_{89}$.
These rotations are part of the Lorentz group in
ten dimensions and, therefore, the four dimensional
supercharges are charged under both $U(1)_{45}$ and
$U(1)_{89}$; hence, the two $U(1)$ symmetries are
R-symmetries.

Summarizing, the charge assignments under
$U(1)_{45}\times U(1)_{89}$ are:
$Q$ and $\tilde Q$ have charges $(0,1)$, the
mass parameters $m$ in \wmq\ have charges $(2,0)$,
and the superspace coordinates $\theta_\alpha$
have charges $(1,1)$.
With these assignments, the mass term \wmq\
is invariant under both global
symmetries\foot{These global symmetries may
be used to argue that the three adjoint
superfields corresponding to fluctuations of
the fourbranes in $(x^4,\cdots, x^9)$ are
infinitely massive; this will be discussed
in section 7.}.

\medskip
\subsec{The magnetic theory}

The magnetic brane configuration is similar to
that discussed in section {\it 4.5}. It contains
$N_c$ fourbranes connecting the \nsp\ fivebrane
to an NS fivebrane on its right (we will refer
to these as fourbranes of the first kind), and
$N_f$ fourbranes (of the second kind) connecting
it to $N_f$  D sixbranes on its left.
We will consider the case $N_f\ge N_c$ in
what follows.

This configuration describes SQCD with gauge
group $U(N_c)$ (with the gauge bosons
coming as before from 4 -- 4 strings connecting
fourbranes of the first kind), $N_f$ flavors of quarks
$q_i$, $\tilde q^i$ (4 -- 4 strings connecting the
$N_c$ fourbranes of the first kind with the $N_f$
fourbranes of the second kind),
and a complex meson $M^i_j$ $(i,j=1,\cdots, N_f)$
arising from 4 -- 4 strings connecting fourbranes
of the second kind. The standard open string coupling
gives rise to a superpotential
\eqn\wmag{W_{\rm mag}=M^i_j q_i\tilde q^j}

The analysis of moduli space and deformations
of this model is similar to the electric theory,
with a few differences due to the existence of
the superpotential \wmag. Consider first mass
deformations. In gauge theory we can add a
mass term to the magnetic quarks, by modifying
the superpotential to:
\eqn\wmagmm{W_{\rm mag}=
M^i_j q_i\tilde q^j+\delta M^i_j q_i\tilde q^j}
The mass parameters $\delta M$ can be absorbed into
the expectation value of the magnetic meson
$M^i_j$, and should thus be thought of as
parametrizing a moduli space of vacua.
The $N_f^2$ resulting parameters are described
in the brane language by splitting the $N_f$ D
fourbranes of the second kind on the D sixbranes
in the most general way consistent
with the geometry. It is easy to check that there
are generically a total of $N_f^2$ massless modes
in this brane configuration, corresponding to the
$N_f^2$ components of $M$: $N_f$ of them describe
fluctuations in the $(x^8, x^9)$
plane of fourbranes stretched between the \nsp\
fivebrane and the rightmost D sixbrane, and
$\sum_{l=1}^{N_f-1} 2l = N_f(N_f-1)$
parametrize fluctuations in
$(x^6,x^7, x^8, x^9)$ of the fourbranes connecting
different sixbranes.

To give the magnetic meson field a mass in gauge theory,
we add a linear term in $M$ to the magnetic superpotential:
\eqn\mwmag{W_{\rm mag}=M^i_j (q_i\tilde q^j-m_i^j)}
Integrating out the massive field $M$ we find that
in the presence of the mass parameters $m_i^j$
the gauge group is broken; thus the
``mass parameters'' $m$ play the role of
Higgs expectation values. In the brane description,
these deformations correspond to a process where
fourbranes of the first kind are aligned with those
of the second kind and reconnected to stretch between
the NS fivebrane and a D sixbrane. If $m$ has rank
$n(\le N_c)$, $n$ such fourbranes are reconnected. The
D sixbranes on which the reconnected fourbranes end
can then be moved in the $(x^4, x^5)$ directions,
taking the fourbranes with them and breaking the
$U(N_c)$ gauge group to $U(N_c-n)$.
The brane description realizes only a subset of the
allowed ``mass'' matrices $m$, namely those which
satisfy \mmdag\ (the reason is similar to the one
described there). We will soon see that this analogy
is not coincidental.

Another deformation of the magnetic gauge theory
and of the corresponding brane configuration, which
will play a role in the sequel, is switching on
a FI D-term for the $U(1)$ subgroup of $U(N_c)$.
In the brane construction this corresponds to a
relative displacement of the NS and \nsp\ fivebranes
in the $x^7$ direction. To preserve SUSY, all
$N_c$ fourbranes of the first kind have to be
reconnected to $N_c$ of the $N_f$ fourbranes of
the second kind, leading to a situation where
$N_c$ fourbranes stretch between the NS fivebrane
and $N_c$ different sixbranes and $N_f-N_c$
fourbranes stretch between the \nsp\ fivebrane
and the remaining sixbranes. Once this occurs,
the two fivebranes can be separated in $x^7$.

Unlike the electric theory, here there
is classically a jump in the dimension of the moduli
space of the theory as we vary the D-term.
For non-vanishing D-term
there are only $N_f-N_c$ fourbranes that give rise
to moduli (the other $N_c$ are frozen because
of the s-rule), and the moduli space is easily checked
to be $N_f^2-N_c^2$ dimensional. When the D-term
vanishes, the previously frozen fourbranes can be
reconnected to yield the original configuration, with
unbroken $U(N_c)$, and we gain access to the full
$N_f^2$ dimensional moduli space discussed above.
Quantum mechanically, this classical jump in the
structure of the moduli space disappears, as will
be discussed in section 6.

The magnetic brane configuration is invariant
under the same global symmetries as the electric
theory described above. It is not difficult to
show that the charge assignments under the
$U(1)_{45}\times U(1)_{89}$ symmetry are as follows:
the magnetic quarks $q$, $\tilde q$ have charges
$(1,0)$, the mass parameters $m$  have charges
$(2,0)$, the magnetic meson $M$ has charges
$(0,2)$, and the superspace coordinates $\theta_\alpha$
have charges $(1,1)$.

\medskip
\subsec{Seiberg's duality in the classical brane picture}

We have now constructed using branes
two $N=1$ supersymmetric gauge theories, the electric
and magnetic theories discussed in the previous
two subsections. Seiberg has shown \nati\ that
the electric gauge theory with gauge group $U(N_c)$ and
the magnetic theory with gauge group $U(N_f-N_c)$
are equivalent in the extreme infrared\foot{Seiberg
actually considered the $SU(N_c)$ and $SU(N_f-N_c)$
theories, but the statement for $U(N_c)$,
$U(N_f-N_c)$ follows from his results by
gauging baryon number.} (\ie\ they flow to
the same infrared fixed point). Seiberg's duality
is a quantum symmetry, but it has classical consequences
in situations where the gauge symmetry is completely
broken and there is no strong infrared dynamics. In
such situations Seiberg's duality reduces to a
classical equivalence of Higgs branches and their
deformations.

In this subsection we show using brane theory
that the Higgs branches of the electric and magnetic
theories with gauge groups $U(N_c)$ and $U(N_f-N_c)$,
respectively, provide different parametrizations
of a single space, the moduli space of vacua of the
appropriate brane configuration. This explains the
classical part of Seiberg's duality. As one approaches
the root of the Higgs branch, non-trivial quantum gauge
dynamics appears, and we have to face the resulting
strong coupling problem. This will be addressed in
section 6.

Start, for example, with the electric theory with gauge
group $U(N_c)$. As we learned above, higgsing
corresponds to breaking D fourbranes on D
sixbranes. It is convenient before entering the Higgs
branch to move the NS fivebrane to the right through
the $N_f$ D sixbranes. This is a smooth process which
does not influence the low energy physics on the
fourbranes \hw. When the NS fivebrane passes through the
$N_f$ D sixbranes, it generates $N_f$ D fourbranes
connecting it to the D sixbranes; these new D fourbranes
are rigid.

We now enter the Higgs phase by reconnecting the $N_c$
original fourbranes to $N_c$ of the $N_f$
new fourbranes created previously; we then further
reconnect the resulting fourbranes in the most general
way consistent with the rules described in section 3.
The resulting moduli space is $2N_fN_c-N_c^2$ dimensional,
as described in section {\it 5.1}. Note that, generically,
there are now $N_f-N_c$ D fourbranes attached to the NS
fivebranes, and $N_c$ D fourbranes connected to the \nsp\
fivebranes (the other ends of all these fourbranes lie on
different D sixbranes).

Once we are in the Higgs phase, we can freely move the
NS fivebrane relative to the \nsp\ fivebrane, and in
particular the two branes can pass each other in the
$x^6$ direction without ever meeting in space. This
can be achieved by taking the NS fivebrane around the
\nsp\ fivebrane in the $x^7$ direction, \ie\ turning
on a FI D-term in the worldvolume gauge theory. At a
generic point in the Higgs branch of the electric
theory, turning on such a D-term is a completely smooth
procedure; this is particularly clear from the brane
description, where in the absence of D fourbranes
connecting the NS fivebrane to the \nsp\ fivebrane,
the relative displacement of the two in the $x^7$
direction can be varied freely.

After exchanging the NS and
\nsp\ fivebranes, the brane configuration
we find can be interpreted as
describing the Higgs phase of {\it
another} gauge theory. To find out
what that theory is, we approach
the root of the Higgs branch by
aligning the $N_f-N_c$ D fourbranes
emanating from the NS fivebrane
with the \nsp\ fivebrane, and the
$N_c$ D fourbranes emanating from
the \nsp\ fivebrane with
D fourbranes stretched between D sixbranes.

We then reconnect the D fourbranes to obtain a
configuration consisting of $N_f-N_c$ D fourbranes
connecting the \nsp\ fivebrane to an NS fivebrane
which is to the left of it; the \nsp\ fivebrane
is further connected by $N_f$ D fourbranes to the
$N_f$ D sixbranes which are to the left of it. This
is the magnetic SQCD of section {\it 5.2}, with
gauge group $U(N_f-N_c)$.

To summarize, we have shown that the moduli
space of vacua of the electric SQCD theory
with (completely broken) gauge group $U(N_c)$
and $N_f$ flavors of quarks $Q^i$, $\tilde Q_i$,
and the moduli space of vacua of the magnetic
SQCD model with (broken) gauge group $U(N_f-N_c)$,
can be thought of as providing different
descriptions of a single
moduli space of supersymmetric brane
configurations. One can smoothly interpolate
between them by varying the scale $\Lambda$,
keeping the FI D-term fixed but non-zero.
Since the only role of $\Lambda$ in the low
energy theory  is to normalize the operators
\ks, theories with different values of
$\Lambda$ are equivalent. The electric and
magnetic theories will thus share all features,
such as the structure of the chiral ring (which
can be thought of as the ring of functions
on moduli space), that are independent of the
interpolation parameter $\Lambda$.

The above smooth interpolation relies on the
fact that the gauge symmetry is completely
broken, due to the presence of the FI D-term.
As mentioned above, it is not surprising that
duality appears classically in this situation
since there is no strong infrared gauge dynamics.

The next step is to analyze what happens as the
gauge symmetry is restored when the D-term goes
to zero and we approach the origin of moduli space.
Classically, we find a disagreement. In the electric
theory, we saw in section {\it 5.1} that nothing
special happens when the gauge symmetry is restored.
New massless degrees of freedom appear, but there
are no new branches of the moduli space that we gain
access to.

In the magnetic theory the situation is different.
When we set the FI D-term to zero, we saw in section
{\it 5.2} that a large moduli space of previously
inaccessible vacua becomes available. While the
electric theory has a $2N_fN_c-N_c^2$ dimensional
smooth moduli space, the classical magnetic theory
experiences a jump in the dimension of its moduli
space from $2N_fN_c-N_c^2$ for non-vanishing
FI D-term, to $N_f^2$ when the D-term is zero.
However, in the magnetic theory when the D-term
vanishes the $U(N_f-N_c)$ gauge symmetry is
restored, and to understand what really happens we
must study the quantum dynamics. We will discuss
this in section 6, where we shall see that quantum
mechanically the jump in the
magnetic moduli space disappears, and the quantum
moduli spaces of the electric and magnetic theories
agree.

It is instructive to map the deformations of the
classical electric theory to those of the classical
magnetic one. Turning on masses \wmq\ in the electric
theory corresponds to moving the D sixbranes away
from the D fourbranes (or equivalently from the
\nsp\ fivebrane) in the $(x^4, x^5)$ directions.
As discussed in section {\it 5.2}, in the magnetic
description, the electric mass parameters correspond
to Higgs expectation values  \mwmag.

Turning on expectation values to the electric
quarks, which was described in the brane language
in section {\it 5.1}, corresponds on the
magnetic side to varying the expectation value of
the magnetic meson $M$, \wmagmm. This gives masses to the
magnetic quarks.

The transmutation of masses into Higgs expectation values
and vice versa observed in the brane construction is one
of the hallmarks of Seiberg's duality.

\bigskip
\newsec{Quantum electric and magnetic N=1 SQCD}
\medskip

To complete the demonstration of Seiberg's
duality in brane theory we have to address
the strong coupling quantum effects which
become important near the origin of moduli
space. In this section we will do that,
taking the opportunity to describe some
additional aspects of the quantum physics of
the electric and magnetic theories in three
and four dimensions.

\medskip
\subsec{Quantum effects in the electric theory}

\noindent
{\sl 1) Global symmetries:}

\noindent
In section {\it 5.1} we saw that the
classical electric theory is invariant
under two chiral $R$ symmetries, corresponding
in the brane language to rotations in
the (4,5) and (8,9) planes. In gauge theory
it is well known that quantum mechanically
only one combination of the two $R$
symmetries is preserved, due to chiral anomalies.
This can be seen directly in brane theory.

At finite type IIA coupling $\lambda$ we should
interpret our brane configurations as describing
fivebranes and sixbranes in M-theory. Defining
(as in section {\it 4.1}):
\eqn\defsvw{
\eqalign{
s=& x^6+i x^{10}\cr
v=& x^4+i x^5\cr
w=& x^8+i x^9\cr
}}
the D fourbrane corresponds
``classically'' to an M-theory
fivebrane at $v=w=0$ and it is
extended in $s$, the NS fivebrane
is at $s=w=0$ and is extended
in $v$, while the \nsp\ fivebrane
is at $v=0$, $s=S$ and is extended
in $w$.

Quantum mechanically \ewew\ the
NS and \nsp\ fivebranes are deformed
by the fourbranes ending on them
according to \sfive. Consider the classical
electric configuration corresponding
to $N_f$ D sixbranes connected by D
fourbranes to an NS fivebrane which is to
their right, with the NS fivebrane further
connected by $N_c$ fourbranes to an
\nsp\ fivebrane which is to its right.
Equation \sfive\ implies that this
classical configuration is deformed
quantum mechanically. Far from
the origin of the $(v,w)$ plane
(the location of the fourbranes),
the location of the NS fivebrane in
the $s$ plane is described by
\eqn\ssff{s_5=R_{10}(N_f-N_c)\ln v}
while the location of the \nsp\
fivebrane in the $s$ plane is
described by
\eqn\ssnsp{s_{5^\prime}=R_{10}N_c\ln w}

Thus, while the classical brane configuration
is invariant both under rotations in
the $(x^4, x^5)$ plane, $v\to e^{i\alpha} v$,
and in the $(x^8, x^9)$ plane, $w\to e^{i\beta} w$,
the quantum mechanical configuration breaks both
symmetries. Nevertheless, one combination of the
two $U(1)$ symmetries is preserved, if accompanied
by an appropriate translation in $x^{10}$.
The unbroken $R$ symmetry is the one under which
\eqn\snsnsp{s_5-s_{5^\prime}=R_{10}\ln \left(
w^{-N_c}v^{N_f-N_c}\right)}
is invariant. It is not difficult to check that
if (by definition) the $R$ charge of $\theta$
under this symmetry is one, that of $Q$,
$\tilde Q$ is $B_f=1-N_c/N_f$, in agreement
with the gauge theory answer \refs{\neqone,
\nonerev}.

\noindent
{\sl 2) Moduli space of vacua for $N_f>N_c$:}

In section 4 we described the quantum
moduli space of $N=1$ SQCD with
$N_f\le N_c$ in four dimensions and
its compactification on a circle, comparing
the gauge theory picture of \ahiss\ to that
obtained by using the brane interactions of
section {\it 4.4}. Here we will comment on
the case $N_f>N_c$ and the role of real masses
in the compactified theory.

As is clear
from \ahiss\ and from the discussion of
section 4, the physics of the four dimensional
theory compactified on a circle of finite
radius $R$ is similar to that of the
infinite $R$ one. As $R\to 0$ one often
encounters different phenomena. It will
be convenient, as in section 4, to discuss
the compactified theory, as it helps to geometrize
more of the gauge theory phenomena. We will also
perform the T-duality transformation of sections
{\it 4.3, 4.4} and discuss threebranes at points on 
a circle of radius $R_3=1/R$ rather than fourbranes
wrapped around a circle of radius $R$.

Consider the electric theory on $R^3\times S^1$
with $N_f>N_c$, and vanishing real masses.
We start with the case $R_3=\infty$ in which the
theory describes a three dimensional $N=2$
supersymmetric QCD. The moduli space of
vacua is similar to that of the theory with
$N_f=N_c$ (see section {\it 4.2}).
There are three branches of
moduli space:

\noindent
\item{a)} Higgs branch: $V_+=V_-=0$. The
meson field $M$ gets an expectation value,
generically breaking the $U(N_c)$ gauge
group completely. The Higgs branch is
$2N_fN_c - N_c^2$ dimensional, and corresponds
in the brane picture to breaking all $N_c$
threebranes that are initially stretched
between the NS and \nsp\ fivebranes on
D fivebranes.

\noindent
\item{b)} Mixed Coulomb -- Higgs branch,
where $V_+V_-=0$, but either $V_+$ or
$V_-$ is massless. This branch is realized
by breaking all but one of the threebranes
on the D fivebranes, and allowing the remaining
threebrane to fluctuate in the $x^3$ direction,
either above ($V_+$) or below ($V_-$) the
location of the D fivebranes.
All repulsive interactions between the single
threebrane stretched between the NS and \nsp\
fivebranes and the broken threebranes are
screened, and hence this branch of moduli
space is not lifted. Its dimension is
$(2N_f(N_c-1)-(N_c-1)^2)+1$ with the term in
parenthesis coming from the mesons $M$ and
the $+1$ from $V_\pm$.

\noindent
\item{c)} Another mixed Coulomb -- Higgs
branch is obtained by breaking all but
two of the threebranes on D fivebranes, and
placing one of the two above and the other
below the D fivebranes in $x^3$.
In this branch $V_+$ and $V_-$ are massless
while the meson moduli space, arising from
the $N_c-2$ broken threebranes, is
$2N_f(N_c-2)-(N_c-2)^2$ dimensional. Again, all
repulsive interactions are screened.

\noindent
The repulsion between threebranes stretched
between NS and \nsp\ fivebranes does not allow
us to leave more than two such threebranes
unbroken. Therefore, the branches discussed
above exhaust the possibilities for unlifted
parts of moduli space. For finite $R_3$ the
structure is very similar, except that the
third branch of the three dimensional
moduli space discussed above is lifted,
because the interaction between the two unbroken
threebranes is no longer screened (as in the
case $N_f=N_c$ discussed in section {\it 4.4}).

In the four dimensional limit $R_3\to 0$ the
Coulomb branches degenerate (the threebranes
have no room to move in the $x^3$ direction),
and we are left with the $2N_cN_f-N_c^2$
dimensional Higgs branch familiar from the
classical theory. An important conclusion is
that in the quantum theory, just like in the
classical one, there is no discontinuous jump
in the structure of the moduli space at the root
of the Higgs branch.

\noindent
{\sl 3) Adding real masses for the quarks:}

In addition to complex masses \wmq, which
correspond to the positions of D fivebranes
in the $(x^4, x^5)$ plane, in the compactified
theory one can add ``real masses'' for the
quarks by displacing the D fivebranes in the
$x^3$ direction (see \bho, \ahiss\ for the
gauge theory description of these mass
parameters). When all $N_f$ real masses
are different (\ie\ the D fivebranes are all
at different values of $x^3$, $a_i$, with
$i=1,\cdots, N_f$), the discussion of the
moduli space has to be modified somewhat.

For $1\le N_f<N_c-1$ there is still no vacuum
both for finite and infinite $R_3$,
since there is no way to place the threebranes
such that all the repulsive interactions are
screened.

For $N_f=N_c-1$ and infinite $R_3$ there
is an $N_c$ dimensional Coulomb 
branch\foot{The existence of such a branch 
was observed in gauge theory in \refs{\bho, 
\ahiss}.} which is described in brane theory 
by placing a D fivebrane between every two 
threebranes, thus screening the repulsive 
interactions between the threebranes. The 
Coulomb branch is then parametrized by
the locations of the $N_c$ threebranes
(and the dual of the $3d$ worldvolume
gauge field), $V_i$ ($i=1,\cdots, N_c=N_f+1$), 
satisfying $a_{i-1}<|V_i|< a_i$ (where $a_0=
-\infty$, $a_{N_f+1}=\infty$). The $U(N_c)$ 
gauge group is broken to $U(1)^{N_c}$ everywhere
on this Coulomb branch. The boundaries of the
Coulomb branch, $|V_i|=a_{i-1}, a_i$, are
singular surfaces with no Higgs branches
emanating from them\foot{This seems to disagree
with \bho.} since whenever a threebrane breaks
on a fivebrane in this situation, unscreened
repulsive forces push parts of it to infinity.

For finite $R_3$ the phase discussed above
disappears, since as we have seen a number
of times before, there is now an unscreened
interaction between the first and last
threebranes.

It is easy to repeat the discussion
above for the case that some of the
real masses are equal. The brane picture
predicts a moduli space corresponding
to particular combinations of
Higgs and Coulomb branches uniquely
determined by the rules of section
{\it 4.4}. We will not go through the
details of that analysis here.
It is also not difficult to generalize
the discussion to $N_f\ge N_c$. Even when
all the real masses are different, there are
now possibilities for mixed Coulomb -- Higgs
branches. Their analysis is straightforward
and will be left to the reader.

\medskip
\subsec{Quantum effects in the magnetic theory}

To complete the demonstration of Seiberg's
duality using branes we now turn to a discussion
of quantum effects in the magnetic SQCD model.
We will restrict to the four dimensional case,
$R=\infty$, and will comment on the compactified
theory in section 10.

The classical magnetic configuration is
invariant under $U(1)_{45}\times U(1)_{89}$
just like the electric one. Quantum mechanically
the NS fivebranes are deformed due to the presence
of the ``fourbranes''; equations \ssff, \ssnsp\
which were found for the electric configuration,
are valid for the magnetic one as well.
This guarantees that the charge assignments
of the various fields ($q$, $\tilde q$, $M$)
agree with those found in gauge theory, and with
the electric configuration.

One may interpret, following \ewew, the
form \ssff, \ssnsp\ of the electric
coupling $s_5-s_{5^\prime}$ as describing a
$v$, $w$ dependent electric QCD scale
$\Lambda_e$:
\eqn\scdep{\Lambda_e^{3N_c-N_f}=
\mu^{3N_c-N_f}
e^{-(s_5-s_{5^\prime})} =\mu^{3N_c-N_f}
\left(w^{-N_c}v^{N_f-N_c}\right)^{-R_{10}}}
where $\mu$ is some fixed scale independent
of $v$, $w$. The first equality in
\scdep\ is exact, while the second
is correct asymptotically, when $v, w$ are
large. The fact that in the magnetic
configuration the NS and \nsp\ fivebranes are
reversed implies that the magnetic QCD scale
$\Lambda_m$ has a different dependence
on $v$, $w$:
\eqn\scmag{\Lambda_m^{3\bar N_c-N_f}=
\mu^{3\bar N_c-N_f}
e^{+(s_5-s_{5^\prime})}}
where $\bar N_c\equiv N_f-N_c$.
Equations \scdep, \scmag\ lead to the scale
matching relation:
\eqn\scmatch{\Lambda_e^{3N_c-N_f}
\Lambda_m^{3\bar N_c-N_f}=\mu^{N_f}}
which is familiar from the study of Seiberg's
duality in gauge theory \refs{\nonerev, \ks}.
Strictly speaking, the argument above shows that
while $\Lambda_e$ and $\Lambda_m$ depend on
$v$ and $w$, for large $v$, $w$ the product 
\scmatch\ is constant. To study the constant
and, in particular, its relation to the 
Yukawa coupling in the magnetic superpotential
\nonerev, one has to perform a more detailed
analysis.

Classically, the moduli spaces of the electric
and magnetic gauge theories were found in section
5 to be different. The electric moduli space is
smooth as the FI D-term approaches zero, while in
the magnetic theory there is a jump in the dimension
of moduli space from $2N_fN_c-N_c^2$ for non zero
D-term, to $N_f^2$ for vanishing D-term. The quantum
analysis of the electric theory performed in
the previous subsection implies that quantum
effects do not lead to a qualitative change in the
structure of the electric moduli space.

In the magnetic theory, quantum effects do
lead to a qualitative change in the structure
of the moduli space. The $N_f-N_c$
fourbranes stretched between the NS and \nsp\
fivebranes are attracted to the $N_f$ fourbranes
stretched between the \nsp\ fivebrane and the
sixbranes (see section {\it 4.4}). Hence,
$N_f-N_c$ of the $N_f$ fourbranes of the second
kind align with the fourbranes of the first
kind, and reconnect, giving rise to $N_f-N_c$
fourbranes stretched between the NS fivebrane
and $N_f-N_c$ different sixbrane (in agreement
with the s-rule). The remaining $N_c$ fourbranes
of the second kind give rise to the usual
$2N_cN_f-N_c^2$ moduli.

In gauge theory, the same conclusions follow
from the fact that the classical magnetic
superpotential \wmag\ is corrected quantum
mechanically to \nati:
\eqn\wmagq{
W_{\rm quantum}=M^i_j q_i\tilde q^j +
\Lambda^{3N_c-N_f\over N_c-N_f} (\det M)^{1\over N_f-N_c}}
Physically, the origin of the second term in
\wmagq\ is the fact that when $M$ gets
an expectation value, the magnetic quarks
become massive due to the classical coupling
\wmag. If the rank of $M$ is larger than
$N_c$, the vacuum is destabilized by
non-perturbative SQCD effects.
This reduces the dimension of moduli space to
$N_f^2-(N_f-N_c)^2=2N_fN_c-N_c^2$.
The analysis of the superpotential \wmagq\
leads to the same conclusions \nati.

Thus, quantum effects eliminate the discontinuous
jump in the chiral ring of the magnetic theory
as the FI D-term is tuned to zero.
Quantum mechanically one does not have access to the
full $N_f^2$ dimensional classical moduli space,
but to a $2N_fN_c-N_c^2$ dimensional subspace
thereof -- precisely the subspace that connects smoothly
to the electric moduli space via our construction!

It is now clear that the equivalence of the chiral rings
of the electric and magnetic theories constructed
above extends to the root of the Higgs branch
(for vanishing D-term), since in the quantum theory
there is no discontinuous jump in the structure of
either the electric or magnetic moduli space there.

\bigskip
\newsec{Theories with two adjoints}
\medskip

\subsec{The brane configuration and its interpretation}

In this section we will consider a configuration
of $k$ coincident NS fivebranes connected by $N_c$
D fourbranes to $k^\prime$ coincident \nsp\
fivebranes, with $N_f$ D sixbranes located between
the NS and \nsp\ branes. The discussion of the
previous sections corresponds to the case
$k=k^\prime=1$.

The low energy theory on the fourbranes is in this
case $N=1$ SYM with gauge group $U(N_c)$, $N_f$
fundamental flavors $Q^i$, $\tilde Q_i$, and two
adjoint superfields $X$, $X^\prime$. The classical
superpotential is
\eqn\wxx{W={s_0\over k+1}\Tr X^{k+1}+
{s_0^\prime\over k^\prime +1}
\Tr {X^\prime}^{k^\prime+1}+
\tilde Q^iX^\prime Q_i}
$X$ and $X^\prime$ can be thought of as describing
fluctuations of the fourbranes in the $(x^8, x^9)$
and $(x^4, x^5)$ directions, respectively.
They are massless, but the superpotential \wxx\ implies
that there is still a potential for the corresponding
fluctuations, allowing only infinitesimal deviations
from the vacuum at $X=X^\prime=0$. The couplings
$s_0$, $s_0^\prime$ should be thought of as very
large: $s_0, s_0^\prime\to\infty$. This can be deduced
\eg\ on the basis of the transformation properties
of \wxx\ under the global symmetries $U(1)_{45}$
and $U(1)_{89}$.

Indeed, in addition to the charge
assignments for $Q$, $\tilde Q$, $m$ and $\theta$
discussed in section {\it 5.1}, the charges of the
adjoint fields under $U(1)_{45}\times U(1)_{89}$
are $(0,2)$ for $X$ and $(2,0)$ for $X^\prime$.
Note that the adjoint field $X^\prime$ transforms
in the same way as the quark mass $m$; this is consistent
with the fact that turning on an expectation value
for $X^\prime$ changes $m$, as is clear from the form
of the superpotential \wxx.

We see that with the above charge assignments for the
fields, the last term in \wxx\ is invariant under
the global symmetry, while the first two are not.
This means that the coefficients $s_0$, $s_0^\prime$
must be infinite. This is consistent with the case
$k=k^\prime=1$ considered in sections 5, 6.
There, the statement is that the masses of $X$ and
$X^\prime$ are not of order $1/L_6$ as one would naively
expect, but rather infinite. That this must be the
case follows from the following argument. If the mass 
of $X^\prime$, $m^\prime$, was finite, by integrating
$X^\prime$ out using \wxx\ we would have obtained a
superpotential of the form
\eqn\supp{W={1\over m^\prime}\tilde Q_i
Q^j\tilde Q_j Q^i}
for the quarks, which would have lifted some or all
of the flat directions of SQCD. The fact that we have
found the full moduli space of SQCD in the brane
construction of section 5 is further evidence that
the mass in that case, or more generally
$s_0^\prime$ in \wxx, is infinite.

One can discuss the case of finite mass for $X^\prime$
by rotating the \nsp\ fivebrane in the $(v,w)$ plane
(as in \barbon). The mass $m^\prime$ varies from zero
when the brane lies along the $v$ axis, to infinity
when it lies along the $w$ axis. In between, there is
a non vanishing superpotential of the form \supp\ for
the quarks, which modifies the moduli space of vacua
rather significantly. One can repeat the counting leading
to \higgsm\ for this case and find that the dimension
of the Higgs branch is $2N_fN_c-2N_c^2$. The counting
\higgsm\ is modified because a) fourbranes stretched
between a D sixbrane and the rotated \nsp\ fivebrane are
now rigid, and b) there is an analog of the s-rule for
fourbranes stretched between a D sixbrane and the rotated
\nsp\ fivebrane. Thus the structure for any non zero
rotation angle is very similar to that of the $N=2$
supersymmetric case discussed in \hw.

The occurrence of infinite coefficients in the
superpotential seems unsatisfactory. It is quite
possible that the description of the theory given
here is an effective one, arising after integrating
out some additional fields, and the underlying theory
with these fields included does not exhibit any such
singularities. We will leave an elucidation of these
issues for future work.

A calculation similar to
that of section {\it 6.1} leads to the
conclusion that only one combination of the two
$U(1)$ $R$ symmetries described above survives
in the quantum theory.
The bending of the NS and \nsp\ fivebranes
is given for arbitrary $k$ and $k^\prime$
(and again for large $v$, $w$) by:
\eqn\defss{\eqalign{
s_5=&R_{10}(N_f-{N_c\over k})\ln v\cr
s_{5^\prime}=&R_{10}{N_c\over k^\prime}\ln w\cr}}
and the unbroken symmetry is the combination
of $v\to e^{i\alpha} v$,  $w\to e^{i\beta} w$
which preserves $s_5-s_{5^\prime}$.

It is not difficult to compute the $R$-charge
of $Q$, $X$ and $X^\prime$ under this symmetry.
One finds that for generic $k, k^\prime$
the answer one gets is not the one expected
from gauge theory (the gauge theory answer for
$k^\prime=1$ and arbitrary $k$ appears in \ks;
for $k^\prime>1$ the gauge theory predicts that
there is no unbroken $R$ symmetry except at
specific values of $N_f/N_c$).

The origin of the discrepancy is clear. The gauge
theory analysis assumes finite $s_0$, $s_0^\prime$
which fixes already classically the $R$ charge of $X$,
$X^\prime$ and $Q$. The brane construction
corresponds to infinite $s_0$, $s_0^\prime$,
in which case the corresponding classical gauge
theory is invariant under {\it three} $U(1)$
symmetries, only two of which are visible in the
brane picture as rotations in ten dimensions.
Therefore, in gauge theory there are in this case
two non-anomalous $U(1)_R$ symmetries, only one
combination of which is seen in the brane picture.

\medskip
\subsec{Deformations and moduli space}

The simplest way to see that the configuration of
branes constructed in the previous subsection indeed
describes a gauge theory with the stated matter
content is to match the deformations of the brane
configuration with those of the gauge theory \wxx.
We start with the parameters corresponding to
the locations of the NS and \nsp\ fivebranes.

Initially, the $k$ NS fivebranes reside
at the same point in the $(x^8, x^9)$ plane.
Displacing them to $k$ different points $a_j=x^8_j+i
x^9_j$, $j=1, \cdots, k$ gives rise to many possible
configurations, labeled by a set of non-negative
integers $(r_1, \cdots, r_k)$, with $\sum_j r_j=N_c$.
The integers $r_j$ specify the number of fourbranes
stretched between the $j$'th NS fivebrane, located at
$a_j$, and the \nsp\ fivebranes which we still take
to be coincident.
Since the $N_c$ fourbranes end on the NS fivebranes,
their locations in $(x^8, x^9)$ follow those of the
NS branes. On the worldvolume of the fourbranes one
can think of the locations $\{a_j\}$ as determining the
expectation value of the adjoint field $X$ describing
fluctuations of the fourbranes in the $(x^8, x^9)$
directions. It is thus clear how the parameters
$a_j$ appear in the worldvolume gauge theory; there
is a superpotential for the adjoint field $X$ of the
form:
\eqn\wadj{W=\sum_{j=0}^k{s_j\over k+1-j} {\rm Tr} X^{k+1-j}}
For generic $\{s_j\}$ the superpotential has $k$
distinct minima $\{a_j\}$ related to the parameters in
the superpotential via the relation:
\eqn\wprime{W^\prime(x)=\sum_{j=0}^k s_j x^{k-j}\equiv
s_0\prod_{j=1}^k(x-a_j)}
Vacua are labeled by sequences of integers
$(r_1, \cdots, r_k)$, where $r_l$ is the
number of eigenvalues of the matrix $X$
residing in the $l$'th minimum of the
potential $V=|W^\prime(x)|^2$. Thus, the
set of $\{r_j\}$ and $\{a_j\}$ determines
the expectation value of the adjoint field $X$,
in agreement with the brane picture. When all
$\{a_j\}$ are distinct, the adjoint field is
massive and the gauge group is broken:
\eqn\brokk{U(N_c)\to U(r_1)\times U(r_2)
\times\cdots\times U(r_k)}
The theory splits in the infrared into $k$ decoupled
copies of SQCD with gauge groups $\{U(r_i)\}$ and
$N_f$ flavors of quarks. The brane description makes
this structure manifest.

The above discussion can be repeated for the
parameters corresponding to the locations of
the $k^\prime$ \nsp\ fivebranes in the $(x^4, x^5)$
directions. These $k^\prime$ complex numbers can
be thought of as parametrizing the extrema of a
polynomial superpotential in $X^\prime$ of order 
$k^\prime+1$, in complete analogy to \wadj, \wprime. 
The only new element is that when we displace the 
$k^\prime$ \nsp\ fivebranes in the $(x^4, x^5)$ 
directions leaving the $N_f$ D sixbranes fixed, 
we make the quarks $Q$, $\tilde Q$ massive, with 
masses of order $\langle X^\prime\rangle$. This 
is the origin of the Yukawa coupling  in the 
superpotential (the last term on the r.h.s. of 
\wxx).

One can also consider situations where both NS and \nsp\
fivebranes are displaced in the $(8,9)$ and $(4,5)$
directions, respectively. There are then $k\times k^\prime$
minima of the potential corresponding to all possible
combinations of the $k$ minima of the $X$ superpotential
and the $k^\prime$ minima of the $X^\prime$ superpotential.
Vacua are labeled by integers $r_{l,m}$ ($l=1,\cdots, k$,
$m=1,\cdots, k^\prime$) specifying the number of eigenvalues
of $X$ and $X^\prime$ that are in the $l$'th minimum of the
potential for $X$ and in the $m$'th minimum of the potential
for $X^\prime$ (with $Q=\tilde Q=0$). Clearly,
$\sum_{l,m}r_{l,m}=N_c$. This is again in agreement with the
brane picture, where $r_{l,m}$ is the number of fourbranes
stretched between the $l$'th NS fivebrane and the $m$'th
\nsp\ fivebrane.

Note that in analyzing the vacuum structure for
$k, k^\prime>1$ we have assumed that we can
diagonalize the matrices $X$ and $X^\prime$
simultaneously. The theory seems to inherit
this property from the $N=4$ supersymmetric
theory on infinite fourbranes. There it can
be understood in gauge theory as arising from
a superpotential of the form $W={\rm Tr}
\left(X^{\prime\prime} [X, X^\prime]\right)$.
Here, the field $X^{\prime\prime}$ is infinitely
massive and the mechanism for enforcing
$[X, X^\prime]=0$ is less clear.

As another check of the brane theory -- gauge theory
correspondence, consider the Higgs moduli space of the
theory. In the brane language, the counting is the
following. Suppose
\eqn\nc{N_c=km+l}
where $m,l$ are non-negative integers,
and $0\le l<k$. As we have seen before, we have to study
all possible ways to decompose the $N_c$ fourbranes into segments
by splitting them on the D sixbranes, taking into account
the s-rule. Since there are now $k$ NS fivebranes, the s-rule
allows $k$ D fourbranes to stretch between them and any
given D sixbrane. Taking this into account and repeating the
analysis done for SQCD above, we find that the dimension of
moduli space is
\eqn\dimh{k\sum_{j=1}^m\left[2(N_f-j)+1\right]+l
\left[2(N_f-m-1)+1\right]=2N_fN_c-km^2-l(2m+1)}
In particular, it is independent of $k^\prime$.

To understand the counting directly
in gauge theory it is convenient \ks\
to slightly deform the singularity
\wxx\ as in \wadj\ (for both $X$ and
$X^\prime$). For generic couplings and
choices of vacua (\ie\ the occupation
numbers $r_{l,m}$ defined previously) the
theory describes $k\times k^\prime$
decoupled SQCD systems, however,
because of the $\tilde Q X^\prime Q$
coupling in \wxx\ at most $k$ of them
contain massless quarks. Thus, the
moduli space we are looking for is that
of $k$ decoupled SQCD systems with gauge
groups $U(r_j)$, $j=1,\cdots, k$ and $N_f$
flavors of quarks. We know from subsection
{\it 5.1} that the dimension of this moduli
space is:
\eqn\modsp{\sum_{j=1}^k
\left(2N_f r_j-r_j^2\right)=2N_fN_c-
\sum_{j=1}^k r_j^2}
We should choose the degeneracies $r_j$
to maximize \modsp. It is easy to see that
the largest dimension is obtained by
taking $l$ of the $r_j$ to be equal to
$m+1$, and the remaining $k-l$ to be equal
to $m$ ($l$ and $m$ are defined by \nc).
Substituting in \modsp, one finds that the
dimension is equal to that deduced from the
brane counting \dimh.

\medskip
\subsec{The magnetic theory and duality for
$k>1$, $k^\prime=1$}

The magnetic brane configuration consists
in this case of an \nsp\ fivebrane connected
by $\bar N_c=kN_f-N_c$ fourbranes to a cluster
of $k$ coincident NS fivebranes on its right, and
by $k$ fourbranes to each of $N_f$ D sixbranes
on its left. The low energy theory on the
fourbranes stretched between fivebranes is
a $U(\bar N_c)$ gauge theory with an adjoint
superfield $Y$ coming from 4 -- 4 strings connecting
fourbranes of the first kind, $N_f$ fundamental
flavors of quarks $q_i$, $\tilde q^i$ ($i=1,\cdots,
N_f$) coming from 4 -- 4 strings connecting fourbranes
of the first kind to those of the second kind, and
$k$ magnetic meson fields $M_j$ ($j=1,\cdots, k$),
each of which is an $N_f\times N_f$ matrix, coming
from 4 -- 4 strings connecting fourbranes of the
second kind\foot{There seem to be
$kN_f\times kN_f$ strings of this sort at the origin
of moduli space, corresponding to $(kN_f)^2$ massless
fields, but this is misleading. At a generic point in
moduli space we will see momentarily that there are
$kN_f^2$ massless meson fields $M_j$.}.

This theory has been shown in \ks\ to be dual to
the electric theory described above. Its
superpotential is
\eqn\wwwmmm{W_m={\bar s_0\over k+1}{\rm Tr} Y^{k+1}
+\sum_{j=1}^kM_j\tilde q Y^{k-j} q}
where $\bar s_0=-s_0$,  and we have set a scale
to one. The analysis of deformations is similar
to that described for $k=1$ in section {\it 5.2}
and will not be repeated for this case.
The classical moduli space is $kN_f^2$ dimensional.
In the gauge theory it corresponds to setting
$q=\tilde q=Y=0$ and turning on arbitrary expectation
values for the singlets $M_j$. In the brane
configuration, since there are $k$ fourbranes
connecting the \nsp\ fivebrane to each of the
$N_f$ sixbranes, one simply gets $k$ copies of the
moduli space of the magnetic theory described in
section {\it 5.2}.

It is easy to generalize the discussion of
duality in sections 5, 6 to $k>1$. By turning on a
relative separation between the $k$ NS fivebranes
and the \nsp\ fivebrane, which again
corresponds to a FI D-term for $U(1)\subset U(N_c)$,
one enters the Higgs phase in both the electric and
magnetic theories, with the gauge group completely
broken. The electric and magnetic Higgs branches,
both of whose dimensions are given by \dimh,
are then seen to parametrize the same space;
as in section 5, they are related by a smooth change
in the scale of the theory, and are therefore
identified. Duality relates the theory with $N_c$
colors to one with $kN_f-N_c$ colors essentially
because the electric configuration contains $k$
fivebranes which are not parallel to the D sixbranes.

As one approaches the origin of the Higgs branch
by switching off the D-term, classical physics is
smooth in the electric theory, while in the magnetic
one the dimension of moduli space jumps from \dimh\
to $kN_f^2$. A straightforward extension of the
analysis of section 6 reveals that quantum mechanically,
due to the attraction between fourbranes of the first
and second kinds one does not have access to the full
$kN_f^2$ dimensional classical moduli space, but only
to the subspace whose dimension is given by \dimh\ that
connects smoothly to the electric description. This
establishes the equivalence between the two theories.

\medskip
\subsec{The magnetic theory and duality for
$k=1$, $k^\prime>1$}

The magnetic brane configuration consists
in this case of $k^\prime$ \nsp\ fivebranes
connected by $\bar N_c=N_f-N_c$ fourbranes
to an NS fivebrane on their right, and
by a single fourbrane to each of $N_f$ D sixbranes
on their left. The low energy theory on the
fourbranes stretched between fivebranes has
a $U(\bar N_c)$ gauge group and an adjoint
superfield $Y^\prime$ coming from 4 -- 4 strings
connecting fourbranes of the first kind, and
$N_f$ fundamental flavors of quarks $q_i$,
$\tilde q^i$ ($i=1,\cdots, N_f$) coming from
4 -- 4 strings connecting fourbranes
of the first kind to those of the second kind.

The adjoint field and the quarks are further coupled
to a single magnetic meson field $M$,
which is an $N_f\times N_f$ matrix, coming
from 4 -- 4 strings connecting fourbranes of the
second kind. The superpotential is:
\eqn\wwwprime{W_m={\bar s_0^\prime\over k^\prime+1}
{\rm Tr} {Y^\prime}^{k^\prime+1}+\tilde qY^\prime q
+M\tilde q q}

The discussion of moduli space of vacua and
deformations of this theory is similar to
previous cases and will be skipped. Its relation
to the electric theory of section {\it 7.1}
is an essentially known result in gauge theory.
It follows from the duality of \ks\ by turning
on a particular perturbation\foot{For $k^\prime=2$
this perturbation has been analyzed in gauge
theory in \ref\asy{O. Aharony, J. Sonnenschein and S.
Yankielowicz, hep-th/9504113, \np{449}{1995}{509}.}.};
for completeness
we review the gauge theory derivation in appendix A.
There is one slight subtlety in the comparison to gauge
theory, which we discuss in the Appendix, having to do
with the question of whether $s_0^\prime$ is finite or
not. For finite $s_0^\prime$, gauge theory predicts that
all magnetic mesons in the dual $U(N_f-N_c)$ gauge theory
are massive. As $s_0^\prime\to\infty$ we show in the
Appendix that one magnetic meson comes down to zero mass,
in agreement with the brane picture.

As explained previously, we can study theories
with finite $s_0^\prime$, by rotating the stack
of \nsp\ fivebranes in the $(v, w)$ plane. The
picture that arises is in complete agreement
with the gauge theory discussion. If the \nsp\
fivebranes are not parallel to the D sixbranes,
the fourbranes of the second kind that stretch
between them are rigid, and hence the magnetic 
meson $M$ becomes massive. In that case, the duality
one finds is just that familiar from gauge theory.

\subsec{The magnetic theory and duality for
$k, k^\prime>1$}

In the general case, the magnetic brane configuration
consists of $k^\prime$ \nsp\ fivebranes connected by
$\bar N_c=kN_f-N_c$ fourbranes to $k$ NS fivebranes on
their right, and by $k$ fourbranes to each of $N_f$ D
sixbranes on their left. The low energy theory on the
fourbranes stretched between fivebranes has
a $U(\bar N_c)$ gauge group and two adjoint
superfields $Y, Y^\prime$ coming from 4 -- 4 strings
connecting fourbranes of the first kind,
$N_f$ fundamental flavors of quarks $q_i$,
$\tilde q^i$ ($i=1,\cdots, N_f$) coming from
4 -- 4 strings connecting fourbranes of the
first kind to those of the second kind, and
$k$ magnetic meson fields $M_j$ ($j=1,\cdots, k$),
each of which is an $N_f\times N_f$ matrix, coming
from 4 -- 4 strings connecting fourbranes of the
second kind.
The superpotential is:
\eqn\wwwprime{W_m={\bar s_0\over k+1}{\rm Tr} Y^{k+1}+
{\bar s_0^\prime\over k^\prime+1}
{\rm Tr} {Y^\prime}^{k^\prime+1}+\tilde qY^\prime q
+\sum_{j=1}^kM_j\tilde q Y^{k-j} q}
In addition, there is a constraint that
enforces $[Y, Y^\prime]=0$, which as in the
electric theory is not fully understood.

This duality with two adjoint fields is not
known in gauge theory. It would clearly be
interesting to complete its analysis; this
is left for future work.

\bigskip
\newsec{Theories with $SO$ and $Sp$ groups: I}

\nref\ejs{N. Evans, C.V. Johnson and A.D. Shapere,
hep-th/9703210.}%

In this and the next section we briefly discuss
the generalization of the results of sections
5 -- 7 to other classical groups. We start in this
section by discussing a configuration of branes
near an orientifold fourplane (which we will denote
by O4); in the next section we will describe
configurations including an orientifold sixplane
(O6). Both constructions give rise to orthogonal
and symplectic groups, depending on the choice
of the orientation projection; in general they differ 
in the field content. Theories with $SO$ and
$Sp$ gauge groups have been also studied recently 
in \refs{\vvv, \ov, \ejs}.

\medskip
\subsec{$SO(N_c)$}

The electric theory corresponds to the following
brane configuration. There is an orientifold plane
O4 with a single \nsp\ fivebrane\foot{As in section
7, one can study the generalization to $k'>1$ \nsp\
fivebranes; we will not consider it here.}
which is stuck at the orientifold (it cannot move in
the $(x^4,x^5,x^7)$ directions because it does not 
have an O4-mirror partner). To the left of the \nsp\ 
fivebrane in the $x^6$ direction there are $k$ NS 
fivebranes and their $k$ mirror images, as well as a
single NS fivebrane which does not have a mirror and, 
therefore, is stuck at the orientifold. The NS 
fivebranes are connected to the \nsp\ fivebrane by a 
total of
\eqn\Nctot{N_c=r_0+2\sum_{j=1}^k r_j}
D fourbranes; $r_0$ fourbranes end on the single NS 
fivebrane which is stuck at the orientifold. $r_j$ 
fourbranes end on the $j$'th NS fivebrane, $j=1,...,k$; 
their $r_j$ mirrors end on the O4-mirror partners of 
the NS fivebranes. We also place $N_f$ D sixbranes 
and their $N_f$ O4-mirror partners between the NS and 
\nsp\ fivebranes. 

The worldvolume dynamics on 
the D fourbranes describes at long distances a
four dimensional $N=1$ supersymmetric gauge theory 
with gauge group $SO(N_c)$, $2N_f$
quarks $Q^i$ in the vector representation, 
and a field $X$ in the adjoint representation. 
The superpotential is
\eqn\sowres{W_e=
\sum_{j=0}^k{s_{2j}\over 2(k+1-j)} \Tr X^{2(k+1-j)}}
As in section 7, all the couplings $\{s_{2j}\}$
should be thought of as tending uniformly to
infinity. Note that: 

\item{1)}
The addition of the O4 plane  
does not change the number of
unbroken supercharges. Therefore, 
our configuration describes at 
low energies a four dimensional 
$N=1$ supersymmetric gauge theory.

\item{2)}
There are ${1\over 2}N_c(N_c-1)$ 
open string sectors connecting 
different D fourbranes and their 
mirrors. These correspond to the
dim$\, SO(N_c)={1\over 2}N_c(N_c-1)$ 
vector multiplets on the worldvolume 
of the D fourbranes. 

\item{3)}
The 4 -- 6 strings connecting the 
$N_f$ D sixbranes and their $N_f$
mirrors to the $N_c$ D fourbranes 
describe $2N_f$ chiral multiplets 
in the fundamental representation 
of $SO(N_c)$. The $Sp(N_f)$ gauge 
symmetry on the D sixbranes 
corresponds to an $Sp(N_f)$ global 
symmetry of the low energy theory 
on the worldvolume of the D 
fourbranes\foot{The maximal flavor 
symmetry of $SO(N_c)$ with $2N_f$ 
quarks  is $SU(2N_f)$; here we only 
see an $Sp(N_f)$ subgroup. This is 
similar to the $U(N_c)$ case, where 
the flavor symmetry seen by the 
brane configuration is the
diagonal $SU(N_f)$ subgroup of the 
$SU(N_f)\times SU(N_f)$ maximal 
global symmetry, as discussed in 
section 5.}.

\item{4)}
As before, a fourbrane stretched between the 
NS and \nsp\ fivebranes can break on D sixbranes 
into pieces with a relative splitting in the
$(x^6,x^7,x^8,x^9)$ directions. This corresponds 
to turning on Higgs expectation values for the 
quarks. The dimension of the Higgs moduli space 
can be obtained, as in section {\it 5.1}, by 
counting the number of all possible breakings. 
For $2k+1=1$ and $N_c=2r$, the result is
\eqn\sobr{\sum_{l=1}^r \left\{2\left[
2N_f-(2l-1)\right]+1\right\}=2N_f(2r)-r(2r-1)}
For $N_c=2r+1$ the result is
\eqn\sobra{\sum_{l=1}^r \left\{
2\left[2N_f-(2l-1)\right]+1\right\}+2(N_f-r)=
2N_f(2r+1)-r(2r+1)}
In \sobr, \sobra\ we use the fact that there are 
two massless 
complex scalars parametrizing fluctuations 
of a D fourbrane stretched between two D 
sixbranes as well as between a D sixbrane 
and its mirror. Moreover, the s-rule allows 
a D fourbrane and its mirror to connect
the NS fivebrane only with a single D sixbrane 
and its mirror. The extra term in \sobra\ is 
due to the D fourbrane which does not have a
mirror partner when $N_c$ is odd. Equations \sobr, 
\sobra\ are in agreement with the gauge theory 
analysis.

\item{5)}
As before, relative positions of the NS and \nsp\
fivebranes in the $x^7$ direction play the role of
FI D-terms in the gauge theory on the fourbranes.

\item{6)}
For generic values of the couplings $\{s_{2j}\}$ 
in \sowres, the bosonic potential 
$V\sim |W^\prime|^2$ has $2k+1$ minima: 
one at the origin, and $k$  paired minima  
at $\{\pm a_j\}$:
\eqn\wppprime{W^\prime(x)=
s_0x\prod_{j=1}^k(x^2-a_j^2)}
If $r_0$ eigenvalues of $X$ sit at zero, and
$r_j$ eigenvalues sit at $\pm a_j$, the gauge
symmetry is spontaneously broken:
\eqn\sonc{SO(N_c)\to SO(r_0)\times
U(r_1)\times \cdots\times U(r_k)}
The only massless matter near the minimum
at the origin is $2N_f$ fundamentals of an 
$SO(r_0)$ gauge group; the non zero minima 
$\{a_j\}$ correspond to SQCD with gauge group 
$U(r_j)$ and $N_f$ quarks in the fundamental 
representation. Thus, for generic $\{a_j\}$, 
the model describes at low energies $1+k$
decoupled supersymmetric QCD theories with 
gauge groups $SO(r_0)$ and $U(r_j)$, 
respectively \sonc. In the brane description, 
$\{a_j\}$ correspond to locations in the 
$(x^8, x^9)$ plane of the $k$ NS fivebranes 
that are free to leave the orientifold plane. 
The parameters $\{r_j\}$ correspond to the
number of fourbranes attached to the different 
NS fivebranes (see \Nctot). Separations of the 
$k$ fivebranes in $(x^6$, $x^7)$ correspond to 
changing the couplings and FI D-terms in the 
$U(r_j)$ factors in \sonc. The rest of the
analysis follows closely that of section 7.

\item{7)}
The distance in the $(x^4,x^5)$ directions between 
the \nsp\ fivebrane and the $N_f$ D sixbranes 
determines the masses of the $2N_f$ chiral multiplets. 
Due to the O4-mirror reflection in the $(x^4,x^5)$ 
directions, the same mass is given to pairs of quarks: 
a quark and its mirror image.

\item{8)}
{\it $R$-charge}:
the inclusion of an orientifold four plane does 
not change the discussion of the classical 
$R$-symmetries presented in section 5. However,
to consider the nonanomalous $R$ symmetry of quarks, 
we need to discuss charges of branes and orientifolds.
An orientifold fourplane has $-1/2$ times the charge 
of a physical D fourbrane (see section 2). Therefore, 
if we normalize the charge of a D fourbrane and its
mirror partner to be $+1$ (so that the total charge of a 
physical D fourbrane is $+2$), the charge of an O4 is $-1$.
When one places NS and/or \nsp\ fivebranes on top of the 
orientifold, the O4 charge flips sign each time it 
crosses an NS or an \nsp\ brane in the $x^6$ direction 
\ejs. Therefore, it looks on the fivebrane as a source 
of charge $-2$, and hence it contributes to the bending 
discussed in section {\it 6.1}. Repeating the analysis 
there one finds that the location of the NS fivebrane in 
the $s$ plane is described by
\eqn\qmns{s_5=R_{10}(2N_f-N_c+2)\ln v}
while the location of the \nsp\ fivebrane is
described by
\eqn\qmnsp{s_{5^{\prime}}=R_{10}(N_c-2)\ln w}
The unbroken $R$-symmetry is the one under which
\eqn\ssnsnsp{s_5-s_{5^\prime}=R_{10}\ln \left(
w^{-N_c+2}v^{2N_f-N_c+2}\right)}
is invariant. It is not difficult to check that
the $R$ charge of $Q$ under this symmetry is 
$B_f=1-(N_c-2)/2N_f$, in agreement with the gauge 
theory answer\foot{In the context of $N=2$ supersymmetric
brane configurations, equation \qmns\ can be used
to generalize the brane calculation of \ewew\ of the
$\beta$ function of $N=2$ SQCD to the other  
classical groups.} \refs{\neqone,\nonerev}.

\noindent  
This concludes the discussion of the electric theory 
and its description in terms of brane dynamics and 
gauge theory.
                            
To find the magnetic theory we follow the discussion
of section {\it 5.3}. Consider first the case $2k+1=1$.
The single NS fivebrane is stuck on the orientifold,
and can no longer avoid crossing the \nsp\ fivebrane
when it approaches it in $x^6$. This is potentially
problematic since to demonstrate Seiberg's duality
on the classical level we want, as in section {\it 5.3},
to embed the electric and magnetic theories in a 
single smooth moduli space of vacua.
   
We propose that the lesson one should draw
from the analysis of section 5 is the following.
The process of moving an NS fivebrane through an
\nsp\ fivebrane is smooth if and only if the 
``linking number'' on each fivebrane is conserved,
without reconnecting branes. 
This linking number \hw\ is defined in the
presence of O4 planes of charge $Q(O4)$ by:
\eqn\lnsofour{l_{NS}={1\over 2}(R_{D6}-L_{D6})+
(L_{D4}-R_{D4})+Q(O4)(L_{O4}-R_{O4})} 
where $L_{D6}$ ($L_{O4}$) [$L_{D4}$] is the number 
of D sixbranes (O4 planes) [D fourbranes] to the 
left of the NS fivebrane and, similarly, $R_{D6}$ 
($R_{O4}$) [$R_{D4}$] is the number of D sixbranes
(O4 planes) [D fourbranes] to its right. 
When the NS fivebrane is to the left of the \nsp\
fivebrane it sees an O4 plane of charge $+1$ on
its left and an O4 plane of charge $-1$ on its right,
therefore, the contribution of the O4 plane to the
linking number is $+2$. Similarly, the contribution
of the O4 plane to the linking number of the \nsp\
fivebrane is $-2$.

To allow a smooth transition one should neutralize 
this difference, by  putting two of the $N_c$ D 
fourbranes on top of the orientifold plane. In 
addition, one must break the remaining $N_c-2$ 
fourbranes on D sixbranes, and enter the Higgs 
branch.  In this situation, there is no obstruction 
to smoothly vary the separation between the two
fivebranes in $x^6$, and one can continue the
discussion as in section 5. 

The magnetic theory, obtained by 
smoothly passing the NS fivebrane 
through the \nsp\ fivebrane and 
approaching the root of the Higgs 
branch, is an $SO(2N_f-N_c+4)$ gauge 
theory with $2N_f$ quarks and a magnetic 
meson field $M$ representing the electric 
bilinears $M=QQ$, coupled to the magnetic 
quarks via a Yukawa interaction.

For general $k$ the analysis is similar. One
can either repeat the previous discussion 
$2k+1$ times\foot{Noting that each time 
an NS fivebrane crosses the \nsp\ fivebrane
the magnetic charge between the remaining 
NS fivebranes and the \nsp\ fivebrane 
flips sign.}, or remove $k$ pairs of NS
fivebranes from the orientifold plane
(breaking the gauge symmetry as in \sonc),
and repeat the discussion of section 5.

After all the $2k+1$ fivebranes cross to
the other side of the \nsp\ fivebrane  
we obtain the magnetic brane configuration, 
in which the $2k+1$ NS fivebranes are connected 
to the \nsp\ fivebrane by $(2k+1)2N_f-N_c+4$ 
fourbranes, and the \nsp\ fivebrane is further 
connected by $(2k+1)\times 2N_f$ D fourbranes 
to the $N_f$ D sixbranes and their $N_f$ mirrors.
This is the magnetic description \ls\ of the 
original theory:

\item{1)} The gauge group on the fourbranes
worldvolume is $G_m=SO(\bar N_c)$, where
\eqn\nct{\bar N_c=(2k+1)2N_f-N_c+4}

\item{2)} The fourbranes connecting the D 
sixbranes to the \nsp\ fivebrane give rise 
to magnetic meson fields $(M_j)_{fg}$, 
$j=0,\cdots,2k$, $f,g=1,\cdots,2N_f$, coming 
from 4 -- 4 string configurations invariant 
under the orientifold projection. $M_j$ is in 
the $N_f(2N_f-1)$ ($N_f(2N_f+1)$) dimensional
antisymmetric (adjoint) representation of the 
$Sp(N_f)$ flavor symmetry for odd (even) $j$.

\item{3)} As in the electric theory, the
locations of the NS fivebranes in the magnetic 
theory are encoded in a polynomial magnetic 
superpotential for an adjoint superfield $Y$, 
corresponding to O4-invariant 4 -- 4 strings
describing fluctuations of the fourbranes in 
the $(x^8,x^9)$ directions.

\item{4)} The couplings of the magnetic
mesons $M_j$ to the magnetic quarks $q$ 
and to the magnetic adjoint $Y$ are as
expected. The full magnetic superpotential 
is:
\eqn\wmsp{W_m= 
\sum_{j=0}^k{\bar s_{2j}\over 2(k+1-j)} \Tr 
Y^{2(k+1-j)}+\sum_{j=0}^{2k}
M_j q Y^{2k-j} q }

\medskip
\subsec{$Sp(N_c)$}

To obtain an $Sp(N_c)$ gauge group instead of 
an $SO(N_c)$ one, all we have to do is to change 
the sign of the orientifold charge\foot{A simple
way to see that D fourbranes near an O4
plane can give rise to both orthogonal and
symplectic gauge groups is to consider two
fivebranes attached to an O4 plane stretched 
around a circle of finite radius in the $x^6$ 
direction. $2N_c$ fourbranes are stretched between 
the two NS fivebranes on one side, and $2N_f$
are stretched on the other. This configuration 
preserves $N=2$ SUSY; if on the $2N_c$ fourbranes 
of the first kind the gauge symmetry is $SO(2N_c)$,
$N=2$ SUSY requires that the global symmetry due to 
the presence of $N_f$ fundamental hypermultiples
is $SP(N_f)$. But this global symmetry is the gauge 
symmetry on the fourbranes of the second kind. This 
implies that the charge of the O4 plane flips sign 
when one passes through an NS (or \nsp) fivebrane. 
A similar sign ambiguity appears for the O6 plane,
to be discussed in section 9; it can be seen by
a simple generalization of the above construction.} 
\refs{\vvv, \ov, \ejs}. Symmetric O4-projections are 
interchanged with antisymmetric O4-projections and, 
therefore, in particular, we are forced to place a total 
even number of D fourbranes: $N_c$ D fourbranes and their 
$N_c$ O4-mirrors. In the presence of $N_f$ D sixbranes 
and their $N_f$ mirrors\foot{An odd number of sixbranes 
leads to a theory with a global anomaly.}, and with
$2k+1$ NS fivebranes, the electric theory is an $N=1$
supersymmetric $Sp(N_c)$ gauge theory with $2N_f$
chiral multiplets in the fundamental representation,
and a field $X$
in the adjoint representation of
$Sp(N_c)$ with a superpotential \sowres.
To obtain the magnetic theory, we follow similar steps 
to subsection {\it 8.1}, and obtain the dual theory with 
superpotential \wmsp. Here we shall present only some of 
the main differences between the $Sp$ and $SO$ projections:

\item{1)}
The dimension of the Higgs moduli space is 
obtained by counting all possible breakings 
of D fourbranes on D sixbranes. For $2k+1=1$ 
the dimension is:
\eqn\sphiggs{\sum_{l=1}^{N_c} \{2\left[2N_f-2l\right]+1\}=
2(2N_f)N_c-N_c(2N_c+1)}
The differences between equations \sphiggs\ and \sobr\ 
are that here the total number of D fourbranes and their 
O4-mirrors is $2N_c$, thus $r$ in \sobr\ is being replaced 
by $N_c$, and the $2l-1$ in \sobr\ is replaced by $2l$
because the antisymmetric projection eliminates one combination
of the motions of fourbranes stretched between sixbranes.

\item{2)}
For generic values of the
$\{a_j\}$ in \wppprime, the bosonic potential
$V\sim |W^\prime|^2$ has $2k+1$ distinct minima: 
one at the origin, and $k$ pairs at $\pm a_j$. 
If $2r_0$ eigenvalues of $X$ sit at the origin, and
$r_j$ 
sit at $\pm a_j$, 
the gauge
symmetry is spontaneously broken:
\eqn\sppnc{Sp(N_c)\to Sp(r_0)\times
U(r_1)\times \cdots\times U(r_k)}
This is clearly seen in the brane description.

\item{3)} {\it R-symmetry:}
following the discussion in sections {\it 6.1, 8.1}, 
we find that quantum
mechanically, the location of the NS fivebrane in the $s$ plane is
described by
\eqn\qmnsss{s_5=2R_{10}(N_f-N_c-1)\ln v}
while the location of the \nsp\ fivebrane in the $s$ plane is
described by
\eqn\qmnssdsp{s_{5^{\prime}}=2R_{10}(N_c+1)\ln w}
The unbroken $R$-symmetry is the one under which
\eqn\snsnssssp{s_5-s_{5^\prime}=-2R_{10}\ln \left(
w^{N_c+1}v^{N_c+1-N_f}\right)}
is invariant. 
The $R$ charge of 
$Q$ is $B_f=1-(N_c+1)/N_f$, in agreement
with the gauge theory answer \ls.

\item{4)} The single NS brane, which is stuck on the
orientifold, can only cross the \nsp\ brane smoothly 
if we neutralize the charge difference between both 
sides of the NS brane and the \nsp\ brane. In the
$Sp$ case, the charge of the part of the orientifold 
between the NS and \nsp\ branes is $+1$, while that 
outside is $-1$. To compensate the difference one can 
do one of two things:
a) modify the electric brane configuration, by replacing
two physical D sixbranes by 
two pairs of semi-infinite fourbranes, one stretching
to the right of the \nsp\ fivebrane, the other to
the left of the NS fivebrane. The charge everywhere
along the orientifold is now $+1$, and the discussion
proceeds as above; b) create a pair of a D fourbrane 
and a D anti-fourbrane together with their mirror 
partners. Now, the two anti-fourbranes can neutralize 
the charge difference along 
the orientifold. After the NS branes cross 
the \nsp\ brane, the anti-fourbranes annihilate with D 
fourbranes, leaving a supersymmetric configuration.
One may regard such anti-fourbranes as virtual.

\noindent
The magnetic theory one finds in this case is
an $Sp(\bar N_c)$ gauge theory with
\eqn\ncctctt{\bar N_c=(2k+1)N_f-N_c-2}
in agreement with \ls.
The matter content and interactions arising from
the brane construction agree with the field theory
studies.

\bigskip
\newsec{Theories with $Sp$ and $SO$ groups: II}

$SO$ and $Sp$ gauge groups with somewhat different
matter content can be constructed using an
orientifold sixplane\foot{It is interesting that
pure SYM theories with gauge groups $Sp(N_c)$ and
$SO(N_c)$ may be obtained from geometrically
different brane configurations, \eg\ ones with
O4 or O6 planes. This may be interpreted as 
an infrared duality in the space of brane theories.}
(O6). Consider an electric brane configuration, 
containing an O6 plane with an \nsp\ fivebrane
embedded in it. To the left of the O6 plane in 
the $x^6$ direction, there are $k$ NS fivebranes
connected to the \nsp\ fivebrane by $N_c$ D 
fourbranes. We also place $N_f$ D sixbranes
between the O6 plane and the stack of NS
fivebranes.

All branes except the \nsp\ fivebrane  
have ``O6-mirrors''. 
In particular, the $k$ NS fivebranes have
$k$ O6-mirror images which can leave the
orientifold plane in pairs. The \nsp\ 
fivebrane cannot be removed from the O6
(and as in section 8, we will not consider 
configurations with more than one \nsp\ 
fivebrane).

The fourbrane worldvolume dynamics describes 
at long distances an $Sp(N_c)$ supersymmetric 
gauge theory with $2N_f$ chiral multiplets in 
the fundamental representation and an antisymmetric 
tensor $X$ in the $N_c(2N_c-1)$ dimensional
representation of $Sp(N_c)$, 
with a superpotential
\eqn\spwres{W_e=
\sum_{j=0}^k{s_{j}\over (k+1-j)} 
\Tr X^{(k+1-j)}}
As in section 8, one notes that:

\item{1)}
The orientifold O6 does not break any additional 
supercharges. Therefore, the configuration has 
four unbroken supercharges, and it describes at
low energies a four dimensional $N=1$ supersymmetric 
gauge theory.

\item{2)}
There are $N_c^2$ 4 -- 4 string sectors 
corresponding to gauge multiplets of
the $U(N_c)$ subgroup of $Sp(N_c)$.
In addition, there are $N_c(N_c+1)$ 
sectors, connecting a fourbrane on one 
side of the orientifold with a fourbrane
on the other side of the orientifold,
which are invariant under the $Z_2$ orientifold 
projection. Altogether, there are $N_c(2N_c+1)$ 
O6-invariant 4 -- 4 string configurations,
corresponding to the dim$\, Sp(N_c)=N_c(2N_c+1)$ 
gauge multiplets on the worldvolume of the D 
fourbranes; the gauge group is $Sp(N_c)$.

\item{3)}
The 4 -- 6 strings streched between the $N_f$ 
D sixbranes and the $N_c$ D fourbranes describe 
$2N_f$ chiral multiplets in the fundamental
representation of $Sp(N_c)$. The $SO(2N_f)$ 
gauge symmetry on the D sixbranes corresponds 
to an $SO(2N_f)$ global symmetry of the low energy
theory on the worldvolume of the D fourbranes.

\item{4)}
As before, it is possible for a fourbrane 
stretched between the NS and \nsp\ fivebranes 
to break on D sixbranes into pieces with a 
relative splitting in the $(x^6,x^7,x^8, x^9)$ 
directions. This corresponds to turning on  
Higgs expectation values for the quarks. To 
obtain the whole Higgs muduli space, we need to 
locate the $N_f$ D sixbranes and their O6-mirrors 
at the orientifold plane. Now, the counting of all 
possible breakings gives for $k=1$:
\eqn\ssphiggs{\sum_{l=1}^{N_c}[4(N_f-l)+1]=
4N_fN_c-N_c(2N_c+1)}
The result \ssphiggs\ agrees with the 
counting in gauge theory.

\item{5)}
One can again describe the deformations
of the superpotential \spwres, $\{s_j\}$,
by displacing NS fivebranes in the  $(x^8, 
x^9)$ plane. For generic values of the 
$\{s_j\}$ the gauge symmetry is spontaneously 
broken: 
\eqn\spnc{Sp(N_c)\to Sp(r_1)\times
Sp(r_2)\times \cdots\times Sp(r_k)}
and the theory corresponding to
the $j$'th minimum is SQCD with gauge group
$Sp(r_j)$ and $2N_f$ fundamentals.

\noindent
The magnetic theory is obtained by going into
the Higgs phase so that no fourbranes connect
the NS fivebrane to the \nsp\ fivebrane, and
then moving the NS fivebrane through the \nsp\
fivebrane in the $x^6$ direction. The only new
element in the discussion is the fact that as
the NS fivebrane passes through the \nsp\ fivebrane
it also meets the O6 plane. 
This process is smooth: as we saw in section 2
the O6 plane behaves like a D sixbrane; its charge
is $-2$ times the charge of a physical sixbrane.

Thus it is reasonable to expect that when the
NS fivebrane passes the O6 plane with an \nsp\
fivebrane embedded in it, two of the fourbranes
connecting it to the D sixbranes reconnect 
to the \nsp\ fivebrane and stay on the 
orientifold. This is consistent with 
conservation of the ``linking number'' 
of \hw\ which reads in the presence of 
orientifolds:
\eqn\lns{l_{NS}={1\over 2}(R_{D6}-L_{D6})+
(L_{O6}-R_{O6})+(L_{D4}-R_{D4})} 
where $L_{D6}$ ($L_{O6}$) [$L_{D4}$] is the 
number of D sixbranes (orientifolds O6) 
[D fourbranes] to the left of the NS 
fivebrane and, similarly, $R_{D6}$ 
($R_{O6}$) [$R_{D4}$] is the number 
of D sixbranes (orientifolds O6) 
[D fourbranes] to its right.

An alternative description of the 
process of moving an NS fivebrane 
through an orientifold which leads 
to the same conclusion is to say 
that since the O6 plane has the 
opposite charge to two D sixbranes,
when an NS brane crosses an O6 plane, 
two D anti-fourbranes are generated 
connecting the NS fivebrane with the 
\nsp\ fivebrane embedded in the 
orientifold six plane. After the NS 
brane passes the orientifold, these 
anti-fourbranes annihilate two 
fourbranes stretched between the NS 
and \nsp\ branes.

After all the $k$ fivebranes cross to 
the other side of the orientifold we 
obtain the following final brane 
configuration. The $k$ NS fivebranes 
are connected to the \nsp\ fivebrane 
by $k(N_f-2)-N_c$ fourbranes. The 
\nsp\ fivebrane is further connected 
by  $k\times N_f$ D fourbranes to the 
$N_f$ D sixbranes. This is the magnetic 
description \intrrr\ of the original theory:

\item{1)} The gauge group on the 
worldvolume of the fourbranes is 
$G_m=Sp(\bar N_c)$, where
\eqn\nnct{\bar N_c=k(N_f-2)-N_c}

\item{2)}
There are $k\times N_f(2N_f-1)$ magnetic mesons
$(M_j)_{fg}$, $j=1,...,k$, $f,g=1,...,2N_f$,
coming as usual from open strings connecting 
fourbranes of the second kind. 

\item{3)} The couplings of the magnetic
mesons $M$ to the magnetic quarks $q$ and 
to the magnetic antisymmetric tensor $Y$ are 
via the magnetic superpotential:
\eqn\wmsp{W_m={\bar s_0\over k+1}\Tr Y^{k+1}+
\sum_{j=1}^k M_j q Y^{k-j} q}

\noindent
Finally, if one flips the charge of the O6 plane
one finds an electric theory corresponding to an
$SO(2N_c)$ gauge theory with $2N_f$ quarks in the 
fundamental representation, and a symmetric tensor 
$X$ with a superpotential \spwres. Duality leads to
a magnetic $SO(2\bar N_c)$ gauge theory, with
\eqn\nctsos{\bar N_c=k(N_f+2)-N_c}
in agreement with the field theory results \intrrr.

\bigskip
\newsec{Comments}
\medskip
\subsec{Quantum brane interactions}

Gauge dynamics and its realization 
in terms of configurations of branes leads
one to deduce that branes that stretch 
between various other branes in general 
interact with each other quantum 
mechanically. In section {\it 4.4} we have 
given a qualitative description of these 
interactions, and noted that they are 
repulsive for branes ending on the same 
side of an \nsp\ fivebrane and attractive
otherwise. One can think of branes 
ending from the left (right) on an
\nsp\ fivebrane as corresponding to
positive (negative) charges. It would be
interesting to understand the precise
form of the interaction between branes
by generalizing the results of \ewew\
to systems with four supercharges, perhaps
as a kind of long range Coulomb 
interaction. 

\medskip
\subsec{Seiberg's duality}

In gauge theory one distinguishes between
two notions of $N=1$ duality. The weaker
version is the statement that members of 
a dual pair share the same quantum chiral 
ring and moduli space of vacua, as a function 
of all possible deformations. In Seiberg's 
original work \nati\ this statement has been 
proven for supersymmetric QCD. 

In generalizations of Seiberg's work (\eg\ 
\refs{\ks-\ils}) it is often difficult to prove 
directly in gauge theory that the full 
chiral rings and moduli spaces of proposed 
duals agree, since to do that it is necessarry
to understand the structure of the classical
moduli space, as well as the full non-perturbative
superpotential on it, both which are in general
difficult tasks. 

The stronger, conjectured, version of duality
asserts that the full infrared limits of 
the electric and magnetic theories coincide. 
The status of the strong duality conjecture
in field theory is not clear. In particular,
in general the chiral ring does not specify
the full infrared conformal field theory.

On a more qualitative level, in field
theory it is not clear {\it why} the 
moduli spaces and chiral rings of members
of a dual pair agree and, consequently, 
constructing duals of complicated theories 
is generally prohibitively complicated. 
The general structure seems quite mysterious.

Embedding the problem in string theory 
sheds light on some of the above issues.
Duality is translated into a statement in
brane mechanics. It is likely that string
theory and brane dynamics serve as a guiding
principle for all gauge theory dualities.

First, it leads to a derivation and better
understanding \egk\ of ``classical $N=1$ 
duality'', which is a consequence of quantum 
duality in situations where the non-abelian 
gauge symmetry is broken and quantum effects 
are weak in the infrared.

One finds that the classical moduli spaces 
of both the electric and magnetic theories 
are embedded in (and provide different
parametrizations of) a single moduli space 
of string vacua. This makes the relation
between the two moduli spaces and chiral 
rings manifest. Furthermore, since both
electric and magnetic gauge groups are broken,
there is no strong infrared dynamics, and the
above equivalence is not corrected quantum
mechanically.

Near the origin of moduli space, the classical 
analysis of the space of brane configurations 
(or, equivalently, moduli space of vacua of the
low energy gauge theories) leads to different
structures on the electric and magnetic 
sides. However, in this situation strong quantum 
infrared dynamics has to be taken into account.
In gauge theory it leads to a dynamically generated
superpotential which lifts part of the classical
moduli space. In string theory, quantum interactions
between branes achieve the same.
After these quantum effects have been taken into
account, the electric and magnetic moduli
spaces, and hence also the quantum chiral rings,
coincide.

The study of field theory duality in 
terms of branes is often more tractable 
than the corresponding gauge theory
analysis. The calculation of quantum 
superpotentials is replaced by a few 
universal rules describing brane 
interactions. This provides a powerful 
tool for analyzing the quantum moduli 
space of vacua. 

The brane construction provides a 
proof of the fact that the quantum 
electric and magnetic moduli spaces 
and chiral rings agree.  It is natural 
to ask whether it also shows the 
equivalence of the full infrared CFT's 
at the origin of moduli space. The 
answer to this question is not known;
the issue is the following.

In theories without exactly marginal
deformations of the infrared conformal 
field theory (in which the IR fixed point 
is an isolated CFT) the leading effect of 
$L_6$, the distance between the NS and 
\nsp\ fivebranes, on the infrared CFT is 
through its influence on the QCD scale 
$\Lambda$, which appears in the low energy
effective action in the gauge field kinetic
term $(\log\Lambda)\int d^2\theta W_\alpha^2$. 
This term is irrelevant at long distances and, 
therefore, long distance physics does not depend 
on $L_6$. However, as the NS fivebrane passes 
the \nsp\ fivebrane through the strong
coupling region at $L_6=0$, it is apriori 
possible that there is a discontinuous jump 
in the physics. It might be that one can 
rule out such a jump by using the fact that
one can go around the singularity by turning 
on a D-term.
   
In theories with exactly marginal
operators, in which there is a
line of infrared fixed points,
changes of $L_6$ will in general
couple to the moduli, and duality
provides a map of the electric
line of fixed points to the magnetic
one. Apart from that, the previous 
discussion of full infrared equivalence  
is similar.

\medskip
\subsec{Other brane configurations}

Among the many possible brane configurations
preserving $N=1$ SUSY in four dimensions that
were not discussed in this paper, an interesting
class of examples is obtained by considering
fourbranes stretched between NS and \nsp\ 
fivebranes in the presence of D sixbranes
parallel to the \nsp\ fivebranes and 
D$^\prime$ sixbranes which are parallel to
the NS fivebranes. 

The brane approach allows the enumeration of
the dimension of the moduli space of vacua
for all possible orderings of D and D$^\prime$
sixbranes along the $x^6$ direction. As an 
example, consider the case of a single fourbrane
stretched between an NS fivebrane and an \nsp\
fivebrane in the presence of $N_f$ D sixbranes
and $N_f^\prime$ D$^\prime$ sixbranes between
the two fivebranes. The dimension of moduli
space depends on the ordering of the sixbranes,
and ranges from $2(N_f+N_f^\prime)-1$ when all
the D$^\prime$ sixbranes are to the left of all
the D sixbranes, to $2(N_f^\prime)-1$ when 
$N_f^\prime>N_f$ and $N_f$ D sixbranes and 
D$^\prime$ sixbranes alternate along the $x^6$
direction. 

It would be interesting to find the corresponding
field theoretic effective potential accounting
for this behavior.

\medskip
\subsec{Other theories}

\nref\ap{I. Antoniadis and B. Pioline,
hep-th/9607058.}
The discussion in terms of branes is clearly
general, and can be applied to other theories.

\medskip
\item{1)} {\it N=2 in d=4}

\noindent
In $N=2$ supersymmetric gauge theories in
four dimensions it can be used to prove
the equivalence of Higgs branches of 
theories with different rank gauge groups, 
\eg\ $U(N_c)$ and $U(N_f-N_c)$ gauge theories 
with $N_f$ flavors of quarks \refs{\ap, \hw}. 
However, as one approaches the root of the 
Higgs branch, it is clear from both the 
classical and quantum brane constructions 
that (for $N_f\not=2N_c$) there is a jump 
in the moduli space and chiral ring (defined 
by choosing an $N=1$ subalgebra). Thus, 
the brane picture predicts that the theories 
at the origin of moduli space are not 
equivalent (unless $N_f=2N_c$).

\medskip
\item{2)} {\it N=2 in d=3} 

\noindent
One can repeat the discussion of sections 
5 -- 9 for $N=2$ supersymmetric gauge 
theories in three dimensions,  
using branes in type IIB string theory. 
For vanishing (or in general equal) real
masses, the brane construction predicts
that the dimensions of the electric and 
magnetic moduli spaces agree, but to 
achieve a more detailed understanding
one needs to find the geometric realization 
of the fields $W_\pm$  introduced in equation
\qmagth. The gauge theory analysis 
of this problem was recently performed
in \ref\aharony{O. Aharony, hep-th/9703215.}
(see also 
\ref\kar{A. Karch, hep-th/9703172.}), where 
it was conjectured that the magnetic theory
described by \qmagth\ is equivalent to
the electric one.  
If one turns on different real masses,
the electric and magnetic theories
cannot agree, since they have different
Coulomb branches (see \ahiss\ and
section 6). Therefore, it is
unlikely that the full electric and
magnetic infrared CFT's agree.
The status of duality in these models
remains to be understood.

\medskip
\item{3)} {\it N=(2,2) in d=2}

\noindent
It would be interesting to study the
implications of the brane construction
in two dimensional gauge theories,
which can be obtained from our four
dimensional construction by T-duality
in (say) $x^2$ and $x^3$. This may
provide a connection to the construction
of \ew.

\nref\bh{J. Brodie and A. Hanany, hep-th/9704043.}%
\nref\bsty{A. Brandhuber, J. Sonnenschein, S. Theisen 
and S. Yankielowicz, hep-th/9704044.}%
\nref\ao{C. Ahn and K. Oh, hep-th/9704061.}%
\bigskip
\noindent
{\bf Note added:} Related recent work 
appears in \refs{\bh-\ao}.

\bigskip
\noindent{\bf Acknowledgements:}
We thank Y. Oz and especially R. Plesser for 
useful discussions. This work is supported in 
part by the Israel Academy of Sciences and 
Humanities -- Centers of Excellence Program. 
The work of A. G. and E. R. is supported in 
part by BSF -- American-Israel Bi-National 
Science Foundation. S. E., A. G. and E. R. 
thank the Einstein Center at the Weizmann 
Institute for partial support.

\bigskip

\appendix{A}{A gauge theory analysis of the case
$k^\prime>1$}

Consider an $SU(N_c)$ gauge theory with a field $X$
transforming in the adjoint representation and $N_f$
flavors of quarks $Q^i$, $\tilde Q_{\tilde i}$, coupled
via the superpotential\foot{Throughout this appendix
we will replace $k^\prime$ by $k$ to simplify the
notation.}:
\eqn\sp{W={s_0 \over{k+1}} \Tr X^{k+1} +
\sum_{i=1}^{N_f} \lambda_i \tilde Q_i X   Q^i}
In the presence of the superpotential \sp\ the
nonanomalous global symmetry of the model is
$[U(1)]^{N_f}$, all the symmetries being
vectorlike. When all the $\lambda_i $ are equal,
the global symmetry is promoted to $U(N_f)$.

In the absence of the second term in the superpotential,
the theory does not have a vacuum for $N_f<N_c/k$ \ks.
The Yukawa coupling stabilizes the model, ensuring the
existence of a vacuum for any $N_f$.  To see that, deform
the $X$ dependent part of the superpotential \sp\ to:
\eqn\spx{W_x=\sum_{i=0}^{k-1}{s_i\over{k+1-i}} \Tr X^{k+1-i}}
and study the theory for small, non-vanishing $\{s_j\}$,
using the fact \ks\ that there is a vacuum for small
non vanishing $\{s_j\}$ if and only if there is a
vacuum for $s_j=s_0\delta_{j,0}$.
Generically, the bosonic potential $V=|W^\prime|^2$ has $k$
distinct minima $\{a_i\}$,  $i=1,\cdots,k$.
If $r_i$ eigenvalues of $X$ are placed in the minimum at
$a_i$, the gauge symmetry is broken:
\eqn\brr{SU(N_c)\to SU(r_1)\times SU(r_2)\times\cdots\times
SU(r_k)\times[U(1)]^{k-1}}
with $\sum r_i=N_c$.  From \sp\ it is clear that generically
all quarks become massive; the mass of the $i$'th flavor
in the $SU(r_j)$ colour group is $ m_i^{(j)}=\lambda_i a_j$.
The theory reduces in the infrared to a direct product of
SQCD's with massive quarks. Such vacua exist for any number
of flavors.

In order to understand the infrared behaviour of the model
with the superpotential \sp\ we can use duality. The model
with $\lambda_j=0$ has a dual description based on the
gauge group  $SU(k N_f-N_c)$ \ks. We can consider the second
term as a deformation of the electric theory and find its
dual by adding the dual perturbation to the magnetic
superpotential. This leads \ks\ to the deformed magnetic
superpotential:
\eqn\spm{\bar W={\bar s_0 \over{k+1}}\Tr Y^{k+1} +{s_0\over\mu^2}
\sum_{j=1}^k M_j \tilde q Y^{k-j} q +
\sum_{i=1}^{N_f}\lambda_i (M_2)_i^i}
where $\bar s_0=-s_0$, $\mu$ is a scale parameter,
and $M_j$ are gauge singlet mesons representing
$\tilde Q X^{j-1}Q$ in the magnetic theory.
In the presence of the perturbation $\lambda_i$
some of the fields gain a mass and we can integrate
them out. To find the vacua of the deformed theory
we solve the $F$-term conditions:
\eqn\fterm{
\eqalign{\tilde q_{\tilde i} Y^p q^j =&0;\;\;(p\not=k-2)\cr
           {s_0\over\mu^2}q_i Y^{k-2}q^j=&
           -\lambda_i \delta_i^j\cr
           \Tr Y^k=&0\cr}}
In addition one has to satisfy the $D$-term conditions.

Consider first the case when only one of the $\{\lambda_i\}$,
say $\lambda_1$, is different from $0$. The solution of
\fterm\ is then:
\eqn\sola{\eqalign{
\tilde q_\alpha^1=&b\delta_{\alpha,1}\cr
q_1^\alpha=&b\delta^{\alpha,k-1}\cr
Y_\beta^\alpha =&b\delta_{\beta-1}^{\alpha}; \;\;
\beta=2, \cdots ,k-1 \cr
(M_j)_1^1=&0\cr
}}
where $b=[{\lambda_1 \mu^2\over \bar s_0}]^{1\over k}$.
The relative normalizations in \sola\ are fixed by the
D-term conditions. Thus, the magnetic gauge group is
broken to $SU(k N_f-N_c-k+1)$. Expanding the superpotential
\spm\ around the solution \sola\ we find that the
magnetic meson fields $(M_j)_1^a$, $(M_j)_a^1$ with
$a=1,\cdots, N_f$, as well as the fields
$\tilde q_{\alpha}^1$, $q_1^{\alpha}$,
$Y_m^{\alpha}$ for $ m=2,...,k-1$ and
$Y_{\alpha}^n $ for $n=1,2,...,k-2$ become
massive due to the Higgs mechanism.

On the other hand $Y_{\alpha}^{k-1}$ and
$Y_1^{\alpha}$ play a special role:
expanding the $Y^{k+1}$ term in \spm\
around \sola\ gives rise to a term
in the superpotential of the form:
$${3\bar s_0\over{k+1}}b^{k-2} Y_{\alpha}^{k-1}
Y_{\beta}^{\alpha} Y_1^{\beta}$$
which makes it clear that in this
gauge, $Y_{\alpha}^{k-1} $ and $Y_1^{\beta}$
represent a massless magnetic quark with a
superpotential of the type \sp\ with a magnetic
Yukawa coupling,
\eqn\magc{\bar\lambda_1={3\bar s_0\over{k+1}}b^{k-2}}

Repeating the procedure for all $N_f$ flavors we
obtain a magnetic theory with gauge group $SU(N_f-N_c)$,
$N_f$ flavors of magnetic quarks $q, \tilde q $,
and a magnetic superpotential $\bar W$:
\eqn\spmag{\bar W={\bar s_0\over{k+1}} \Tr Y^{k+1} +
\sum_{i=1}^{N_f} \bar\lambda_i
\tilde q_i Y q^i }
where $\bar\lambda_i={3 \bar s_0\over{k+1}}
[{\lambda_i\mu^2\over \bar s_0}]^{k-2}$

Since the leading coefficients of the beta function
in the electric and magnetic theories are
$2 N_c-N_f $ with opposite signs, it is clear
that the electric theory will have the following
three regimes:

\noindent
\item{a)} $N_f\le N_c$: a supersymmetric vacuum of the
theory exists but the IR is trivial all the states
being massive.

\noindent
\item{b)} $N_c< N_f < 2 N_c $: the electric theory is
dual to an infrared free magnetic theory. Therefore, the
massless spectrum is given by the fields appearing in
the magnetic lagrangian.

\noindent
\item{c)} $N_f \ge 2 N_c$: the electric theory is infrared
free and, therefore, the spectrum is given by the fields
appearing in the electric lagrangian.

\noindent
We see that in the process of turning on the Yukawa
couplings $\lambda_i$ in the electric theory, all the
magnetic mesons of \ks\ become massive. At first sight
this seems in contradiction with the fact that in the
case $k=1$ \sp\ reduces to SQCD, where it is known \nati\
that a singlet magnetic meson $M$ is necessary for
duality. The resolution is that $s_0$ in \sp\ is
in that case a mass, so the electric theory is in fact
SQCD perturbed by a quartic superpotential of the form
\supp, generated in the process of integrating out the
massive field $X$. On the magnetic side this translates
to a mass term for the magnetic meson $M$, which can
therefore be integrated out, giving rise to a quartic
superpotential in terms of the magnetic quarks, in
agreement with the direct analysis of \spmag.

Thus, the fact that the equivalence between \sp\
and \spmag\ does not require singlet mesons, relies
crucially on the fact that the parameter $s_0$ is
finite. For infinite $s_0$ in the case $k=1$,
the coefficient of the quartic electric superpotential
\supp\ vanishes and the magnetic meson of SQCD, $M$
becomes massless again.

For $k>1$ one can repeat the same analysis; it is easy
to see by deforming the $X$ dependent part of the
superpotential as in \spx\ and using the previous
result for SCQD that in the limit $s_0\to\infty$
one still needs only one magnetic meson, in agreement
with the result we got from brane dynamics in the text.

\listrefs
\end